\documentclass[11pt]{article}
\IfFileExists{lmodern.sty}{\usepackage{lmodern}\usepackage[T1]{fontenc}}{}
\usepackage[margin=1in]{geometry}
\usepackage{amsmath,amssymb,amsthm,mathtools}
\usepackage{thm-restate}
\usepackage[shortlabels]{enumitem}
\setlist{topsep=5pt,itemsep=3pt}
\usepackage{microtype}
\usepackage{framed}
\usepackage{xcolor}
\usepackage{tikz}
\usetikzlibrary{calc,decorations.pathmorphing}
\usepackage{todonotes}
\usepackage[colorlinks=true,linkcolor=blue!55!black,citecolor=blue!55!black,urlcolor=blue!55!black]{hyperref}
\usepackage[nameinlink,noabbrev,capitalise]{cleveref}

\numberwithin{equation}{section}
\newtheorem{theorem}{Theorem}[section]
\newtheorem{lemma}[theorem]{Lemma}
\newtheorem{corollary}[theorem]{Corollary}

\newtheorem{observation}[theorem]{Observation}

\crefname{hypothesis}{Hypothesis}{Hypotheses}
\theoremstyle{definition}

\newtheorem{remark}[theorem]{Remark}

\newcounter{algorithm}
\crefname{algorithm}{Algorithm}{Algorithms}
\newenvironment{algobox}[1]{\begin{figure}[tbp]\begin{framed}\refstepcounter{algorithm}\noindent\textbf{Algorithm~\thealgorithm}\ (#1)\par\vspace{2pt}\noindent\ignorespaces}{\end{framed}\end{figure}}

\let\oldtextsc\textsc
\renewcommand{\textsc}[1]{\textup{\oldtextsc{#1}}}

\newcommand{\Q}{\mathbb{Q}}

\newcommand{\cU}{\mathcal{U}}
\newcommand{\cT}{\mathcal{T}}

\newcommand{\ex}{\operatorname{ex}}
\newcommand{\Csm}{C^{\mathrm{sm}}}
\DeclareMathOperator{\supp}{supp}

\title{Counting over Unbreakable Decompositions:\\[3pt]
A Single-Exponential Parameterized Algorithm for Minimum Bisection}
\author{Aditya Anand\thanks{Max Planck Institute for Informatics, Saarland Informatics Campus, Germany. \texttt{adanand@mpi-inf.mpg.de}.}
\qquad Evangelos Kipouridis\thanks{Max Planck Institute for Informatics, Saarland Informatics Campus, Germany. \texttt{kipouridis@mpi-inf.mpg.de}.}
\qquad Karol W\k{e}grzycki\thanks{Max Planck Institute for Informatics, Saarland Informatics Campus, Germany. \texttt{kwegrzyc@mpi-inf.mpg.de}. Supported by the Deutsche Forschungsgemeinschaft (DFG, German Research Foundation), grant number 559177164.}}
\date{}

\begin{document}
\hypersetup{pageanchor=false}
\pagenumbering{gobble}
\maketitle

\begin{abstract}
The \textsc{Minimum Bisection} problem asks whether an $n$-vertex graph can be split into two halves with at most $k$ edges between them. We give a single exponential time algorithm for minimum bisection, running in time $2^{O(k)}n^{O(1)}$.
This improves the previous bound of $2^{O(k\log k)}n^{O(1)}$ due to Cygan, Komosa, Lokshtanov, Pilipczuk, Pilipczuk, Saurabh, and Wahlstr\"om (TALG 2021). 

Our algorithm \emph{counts} solutions instead of only finding if one exists. We show that our dynamic program can be implemented on top of an unbreakable decomposition using algebraic techniques.
Along the way, we compute in time $2^{O(k)}n^{O(1)}$ a tree decomposition with adhesions of size at most $2k$ whose bags are $(3k,k)$-edge-unbreakable inside their own subtrees.

\end{abstract}

\newpage
\pagenumbering{arabic}
\hypersetup{pageanchor=true}

\section{Introduction}\label{sec:intro}

Graph partitioning problems ask to cut a graph into parts with few edges between them, subject to a constraint on the parts. When the constraint is local, such as separating given terminals, parameterized algorithms are well understood. Important separators of Marx~\cite{Mar06} and their refinements give fast algorithms for \textsc{Multiway Cut}~\cite{Mar06,CLL09} and for a whole family of related problems~\cite[Chapter~8]{CFKLMPPS15}. Global constraints are less well understood. The most prominent example is \textsc{Minimum Bisection}: given a graph $G$ on $n$ vertices and an integer $k$, can $V(G)$ be split into two sets of sizes $\lfloor n/2\rfloor$ and $\lceil n/2\rceil$ with at most $k$ edges between them?

\textsc{Minimum Bisection} is NP-hard~\cite{GJS76}, and it is one of the classical problems of approximation algorithms. Feige and Krauthgamer gave the first polylogarithmic approximation~\cite{FK02,FK06}, R\"acke obtained an $O(\log n)$-approximation~\cite{Rac08}, and Khot showed that there is no PTAS unless $\mathrm{NP}\subseteq\bigcap_{\varepsilon>0}\mathrm{BPTIME}(2^{n^\varepsilon})$~\cite{Kho06}.

Parameterized by the cut size $k$, the complexity of \textsc{Minimum Bisection} remained open for a long time, and the reason is precisely the global constraint. Recursive understanding, developed by Grohe, Kawarabayashi, Marx, and Wollan~\cite{GKMW11} and by Kawarabayashi and Thorup~\cite{KT11}, and the randomized contractions of Chitnis, Cygan, Hajiaghayi, Pilipczuk, and Pilipczuk~\cite{CCHPP16} solve many cut problems by recursion. They find a part of the graph with a small boundary, solve the problem inside it, and replace it by a small equivalent gadget. What remains is highly connected, in the sense that every cut of size at most $k$ has a small side, and there color coding~\cite{AYZ95} finishes the job. For \textsc{Minimum Bisection} the recursive step fails: whether a part can be replaced depends on the sizes of all other parts. Cygan, Lokshtanov, Pilipczuk, Pilipczuk, and Saurabh~\cite{CLPPS19} circumvented this with \emph{unbreakable tree decompositions}, tree decompositions with small adhesions whose bags are highly connected in the above sense. A dynamic program over such a decomposition handles the global constraint by indexing its tables with sizes. This gave the first fixed-parameter algorithm for \textsc{Minimum Bisection}, with running time $2^{O(k^3)}n^3\log^3n$. Cygan, Komosa, Lokshtanov, Pilipczuk, Pilipczuk, Saurabh, and Wahlstr\"om~\cite{CKLPPSW21} then constructed decompositions with optimal parameters, using lean decompositions, and improved the running time to $2^{O(k\log k)}n^{O(1)}$. Unbreakable decompositions have since become a standard tool, for example for \textsc{Min $k$-Cut}~\cite{LSS20} and \textsc{Fair Bisection}~\cite{ILSS23}, and they can now be even computed in linear time~\cite{ALLLS25,Kor25}.

Two steps of this approach lose a factor of $\log k$ in the exponent. The first is the construction of the decomposition. In their conclusion, Cygan et al.~\cite{CKLPPSW21} ask whether it can be done in time $2^{O(k)}n^{O(1)}$, and note that current techniques fall short of this goal. The second step is the dynamic program itself, and it is the more fundamental one. Inside a bag, a solution puts only $O(k)$ bag vertices on one side, but the bag may have $\Theta(n)$ vertices, so there are $n^{O(k)}$ possibilities for this small side. A dynamic program that minimizes the cut size has to find the right one. Previous algorithms guess its important parts by color coding, as in the randomized contractions framework~\cite{CCHPP16,CKLPPSW21}, and this alone costs $2^{O(k\log k)}$.

\subsection{Our contribution.} All our algorithms are deterministic. We show improved algorithms for both steps and obtain a single-exponential time algorithm for {\textsc{Minimum Bisection}}. Our dynamic program stores \emph{counts} instead of costs: every node of the decomposition counts its partial solutions by size, by cut, and by how they meet the adhesion to its parent. The advantage is that counting supports subtraction and addition, and algebraic ideas can be leveraged. On a high level, we can write a generating function for this counting, so that inside a bag, the logarithm of the generating function splits into
local terms, using a connected-support principle previously used in the setting of cluster expansions and algorithms for graph-polynomial
coefficients~\cite{Bar16,PR17,HPR19,PR19}.

For this to work, we need to construct a faster edge-unbreakable decomposition. Like~\cite{ALLLS25}, the bags are unbreakable not only in $G$ but also in the graph of their own subtree, and we show how to compute such decompositions in time $2^{O(k)}n^{O(1)}$.

We state these results below.

\paragraph{Single exponential time algorithms}\label{sec:intro-counts}

For a graph $G$ and a set $X\subseteq V(G)$, let $\delta_G(X)$ be the set of edges with exactly one endpoint in $X$, and let $|\delta_G(X)|$ be the cut size of $X$. We omit graph subscripts when the graph is clear from the context, writing, for example, $\delta(X)$ instead of $\delta_G(X)$.
A \emph{bisection} of $G$ is a partition of $V(G)$ into two sets of sizes $\lfloor n/2\rfloor$ and $\lceil n/2\rceil$, and its \emph{width} is the number of edges between them. 
Our main result counts the sets of every size with every cut size up to $k$.
\begin{theorem}\label{thm:main}
There is an algorithm that, given a graph $G$ with $n$ vertices and an integer $k\ge0$, computes in time $2^{O(k)}n^{O(1)}$, for all $0\le\ell\le n$ and $0\le j\le k$, the number
\[
C_G(\ell,j)\;=\;\bigl|\{X\subseteq V(G):\ |X|=\ell\text{ and }|\delta(X)|=j\}\bigr| .
\]
\end{theorem}

Applying \cref{thm:main} with $\ell=\lfloor n/2\rfloor$ and checking whether $C_G(\ell,j)>0$ for some $0\le j\le k$ gives the following corollary. 

\begin{corollary}[Minimum Bisection]\label{cor:bisection}
\textsc{Minimum Bisection} can be solved in time $2^{O(k)}n^{O(1)}$. %
\end{corollary}
Thus, we answer the question of Cygan et al.~\cite{CKLPPSW21} about the existence of such an algorithm positively. We prove \cref{thm:main,cor:bisection} in \cref{sec:proof-main}.

\paragraph{Fast unbreakable decompositions}\label{sec:intro-decomp}

One of our technical contributions in proving~\cref{thm:main} is a faster construction of unbreakable decompositions. Let $(\cT,\beta)$ be a rooted tree decomposition of $G$. Following~\cite{CKLPPSW21}, for a node $t$, let $\gamma(t)$ be its \emph{cone}, the union of the bags in the subtree rooted at $t$, let $\sigma(t)\coloneqq\beta(t)\cap\beta(t')$ be its \emph{adhesion}, where $t'$ is the parent of $t$ (and $\sigma(t)\coloneqq\emptyset$ at the root), and let $\alpha(t)\coloneqq\gamma(t)\setminus\sigma(t)$ be its \emph{component}. The \emph{subtree graph} of $t$ is
\[
G_t\coloneqq G[\gamma(t)]-E(G[\sigma(t)]),
\]
where $E(G[X])$ is the set of edges with both endpoints in $X$. Thus $G_t$ is the graph induced by $\gamma(t)$ without the edges inside the adhesion; these edges are accounted for at the parent. A set $\beta\subseteq V(H)$ is \emph{$(q,k)$-edge-unbreakable} in a graph $H$ if every $X\subseteq V(H)$ with at most $k$ edges of $H$ leaving it satisfies $|X\cap\beta|\le q$ or $|\beta\setminus X|\le q$.

Unbreakable decompositions have been constructed with several trade-offs. Cygan et al.~\cite{CLPPS19} compute in time $2^{O(k^2)}n^{O(1)}$ a decomposition with adhesions of size $2^{O(k)}$ whose bags are $(2^{O(k)},k)$-unbreakable in their subtree graphs. The unbreakable decompositions of~\cite{CKLPPSW21} have adhesions of size at most $k$ and bags that are $(i,i)$-unbreakable in $G$ for all $i\le k$, and take time $2^{O(k\log k)}n^{O(1)}$; Korhonen computes lean decompositions (a stronger notion which implies the same guarantees) in time $k^{O(k^2)}n^{O(1)}$~\cite{Kor25}.
Anand, Lee, Li, Long, and Saranurak~\cite{ALLLS25} compute decompositions whose bags are unbreakable in their subtree graphs, with parameters $O(k/\varepsilon)$, in randomized time $2^{O((k/\varepsilon)\log(k/\varepsilon))}n^{O(1)}$. Very recently,~\cite{ALLS27} improves this to a $(5k,k)$ unbreakable decomposition with adhesion $4k$ in time $\tilde{O}(k^{O(k)}m)$. All of these constructions are unbreakable with respect to vertex separations, which implies edge-unbreakability with the same parameters.
Lokshtanov, Saurabh, and Surianarayanan~\cite{LSS20} compute $((k+1)^5,k)$-edge-unbreakable decompositions in polynomial time.

Let $N_G(X)$ be the set of vertices outside $X$ with a neighbor in $X$.
\begin{restatable}{theorem}{decompositiontheorem}\label{thm:decomp}
There is an algorithm that, given a connected graph $G$ with $n$ vertices and an integer $k\ge1$, computes in time $2^{O(k)}n^{O(1)}$ a rooted tree decomposition of $G$ with at most $n$ nodes and the following properties.
\begin{enumerate}[label=(D\arabic*),leftmargin=*]
\item Every adhesion has at most $2k$ vertices.
\item For every non-root node $t$, the component $\alpha(t)$ is nonempty, $G[\alpha(t)]$ is connected, and $N_G(\alpha(t))=\sigma(t)$.
\item Every bag $\beta(t)$ is $(3k,k)$-edge-unbreakable in the subtree graph $G_t$.
\end{enumerate}
\end{restatable}

\begin{figure}[t]
\centering
\begin{tikzpicture}[
  font=\small,
  vtx/.style={circle,fill=black,inner sep=1.4pt},
  tnode/.style={circle,draw=gray!65,fill=white,thick,inner sep=2.2pt},
  bag/.style={draw=red!75!black,very thick,rounded corners=8pt},
  child/.style={draw=orange!80!black,thick,rounded corners=6pt},
  adhesion/.style={draw=orange!70!black,thick,rounded corners=6pt},
  attachment/.style={draw=orange!80!black,thick},
  gedge/.style={draw=gray!85,thin,line cap=round,line join=round},
  lab/.style={inner sep=2pt}]
  \begin{scope}[yshift=0.3cm]
    \path[draw=gray!60,densely dashed,rounded corners=8pt,fill=gray!3]
      (-0.3,-2.5) rectangle (3.3,1.2);
    \draw[gray!65,thick]
      (1.5,2.05)--(1.5,0.65)
      (1.5,0.65)--(0.55,-0.8)
      (1.5,0.65)--(2.45,-0.8)
      (0.55,-0.8)--(0,-1.85)
      (0.55,-0.8)--(1.1,-1.85)
      (2.45,-0.8)--(2.45,-1.85);
    \node[tnode] at (1.5,2.05) {};
    \node[tnode,draw=red!75!black,fill=red!10] at (1.5,0.65) {};
    \foreach \p in {(0.55,-0.8),(2.45,-0.8),(0,-1.85),(1.1,-1.85),(2.45,-1.85)}
      \node[tnode] at \p {};
    \node[lab,anchor=west] at (1.67,2.05) {$t'$};
    \node[lab,red!75!black,anchor=west] at (1.67,0.65) {$t$};
    \node[lab,anchor=east] at (0.38,-0.8) {$c_1$};
    \node[lab,anchor=west] at (2.62,-0.8) {$c_2$};
    \node[lab] at (1.5,2.65) {$\cT$};
  \end{scope}
  \draw[->,gray!65,thick] (3.45,0.95)--(4.25,0.95);

  \begin{scope}[xshift=4.7cm]
    \path[draw=gray!65,densely dashed,rounded corners=10pt,fill=orange!6]
      (-0.25,-2.8) rectangle (9.05,3.1);
    \fill[teal!8,rounded corners=8pt] (0.25,1.7) rectangle (8.55,2.75);
    \draw[bag] (0.25,0.1) rectangle (8.55,2.75);
    \node[lab,anchor=south] at (4.4,3.15) {$\gamma(t)$ (cone)};
    \node[lab,teal!75!black] at (3.5,2.45) {$\sigma(t)=\beta(t)\cap\beta(t')$};
    \node[lab,red!75!black,anchor=east] at (8.35,2.45) {$\beta(t)$};
    \node[lab,orange!70!black] at (4.4,-2.32)
      {$\alpha(t)=\gamma(t)\setminus\sigma(t)$};

    \path[adhesion,fill=orange!22] (0.73,0.58) rectangle (3.32,1.35);
    \path[adhesion,fill=orange!22] (5.4,0.58) rectangle (8.0,1.35);
    \path[child,fill=orange!10] (0.65,-2.05) rectangle (3.4,-0.65);
    \path[child,fill=orange!10] (5.3,-2.05) rectangle (8.1,-0.65);
    \draw[attachment]
      (0.98,0.58)--(0.98,-0.65)
      (3.07,0.58)--(3.07,-0.65)
      (5.65,0.58)--(5.65,-0.65)
      (7.75,0.58)--(7.75,-0.65);
    \node[lab,orange!70!black] at (2.025,-1.78) {$\alpha(c_1)$};
    \node[lab,orange!70!black] at (6.7,-1.78) {$\alpha(c_2)$};

    \draw[gedge]
      (2,1.95) .. controls (3.2,1.95) and (3.8,1.55) .. (4.35,0.78)
      (4.35,1.95)--(4.35,0.78)
      (6.7,1.95) .. controls (5.5,1.95) and (4.9,1.55) .. (4.35,0.78)
      (1.25,0.78)--(2.8,0.78)--(4.35,0.78)--(5.95,0.78)--(7.45,0.78)
      (1.25,0.78)--(1.25,-1.05)--(2.8,-1.05)--(2.8,0.78)
      (5.95,0.78)--(5.95,-1.05)--(7.45,-1.05)--(7.45,0.78);
    \foreach \p in {(2,1.95),(4.35,1.95),(6.7,1.95),
      (1.25,0.78),(2.8,0.78),(4.35,0.78),(5.95,0.78),(7.45,0.78),
      (1.25,-1.05),(2.8,-1.05),(5.95,-1.05),(7.45,-1.05)}
      \node[vtx] at \p {};
    \node[lab,orange!70!black] at (2.025,1.15) {$\sigma(c_1)$};
    \node[lab,orange!70!black] at (6.7,1.15) {$\sigma(c_2)$};
  \end{scope}
\end{tikzpicture}
\caption{The decomposition of \cref{thm:decomp} at a non-root node $t$ with parent $t'$. Red outlines the bag $\beta(t)$; green denotes $\sigma(t)$ and orange denotes $\alpha(t)$. The cone $\gamma(t)$ (gray dashed boundary) is the disjoint union of $\sigma(t)$ and $\alpha(t)$, where $G[\alpha(t)]$ is connected and $N_G(\alpha(t))=\sigma(t)$. Each child component $\alpha(c_i)$ attaches through $\sigma(c_i)\subseteq\beta(t)$. The depicted child adhesions lie in $\alpha(t)$ (darker shading), although in general they may intersect $\sigma(t)$. Every adhesion has at most $2k$ vertices, and $\beta(t)$ is $(3k,k)$-edge-unbreakable in $G_t$. Gray lines in the right panel show representative edges of $G_t$, which omits edges inside $\sigma(t)$.}
\label{fig:decomposition}
\end{figure}
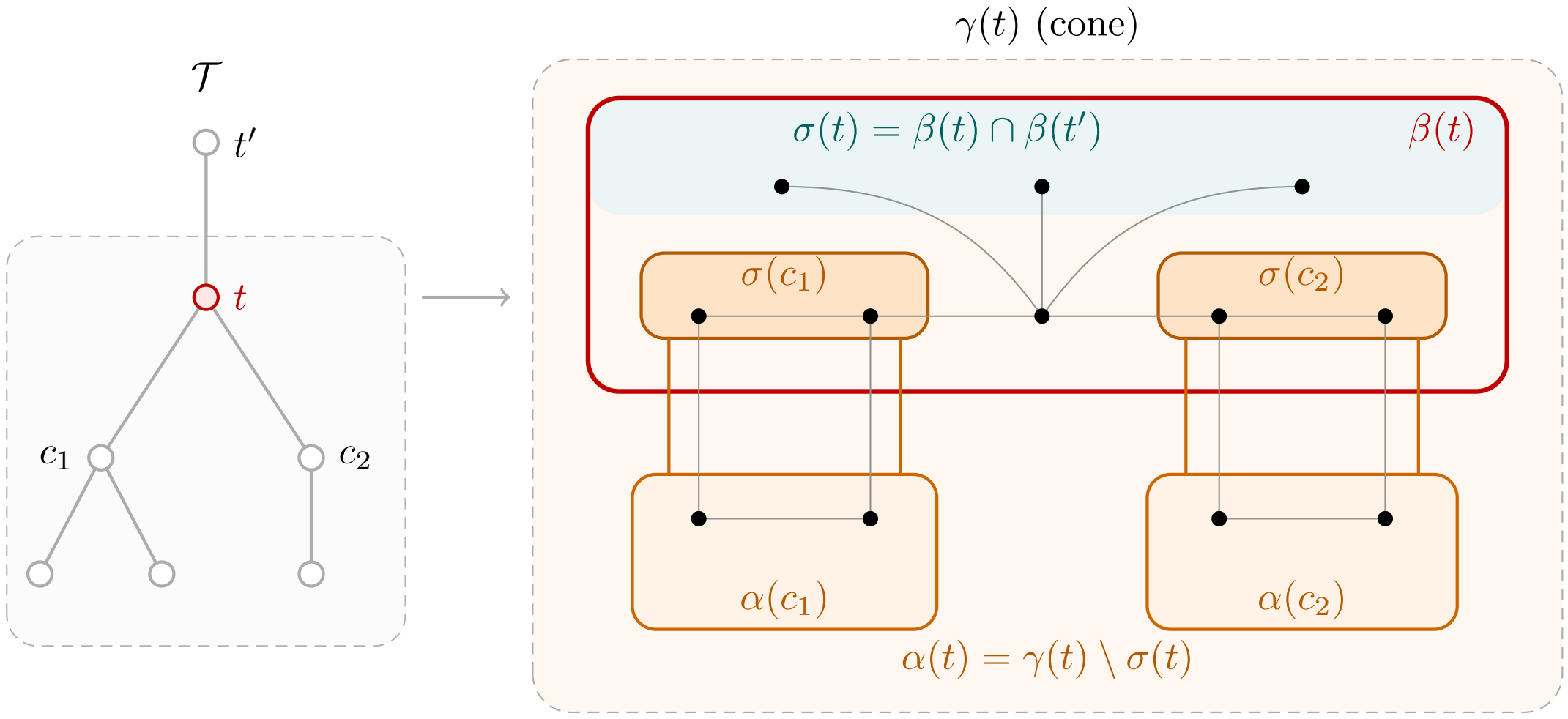

By property (D2) a node is attached to its parent's bag only through its adhesion (see \cref{fig:decomposition}; such decompositions are called \emph{compact}~\cite{CKLPPSW21,ALLLS25}).
Property (D3) is stronger than unbreakability in $G$, since $G_t$ is a subgraph of $G$; it is called subtree unbreakability in~\cite{ALLLS25}, and our dynamic program needs it.
To the best of our knowledge, \cref{thm:decomp} is the first construction in single-exponential time with adhesions of size $O(k)$ and $(O(k),k)$-edge-unbreakable bags. It does not provide their full vertex-unbreakability guarantees, so it does not answer the question of~\cite{CKLPPSW21} in full. But at the scale $k$, its adhesions and its unbreakability are within factors $2$ and $3$ of those of~\cite{CKLPPSW21}, and this suffices for \textsc{Minimum Bisection}. We prove \cref{thm:decomp} in \cref{sec:decomp}.

\paragraph{Other related work.}
Counting and algebraic cancellation are widely used in exact and
parameterized algorithms. For some algorithms based on Inclusion--exclusion, M\"obius transforms,
subset convolution, and their applications to dynamic programming see~\cite[Chapters~10--11]{CFKLMPPS15}.
A prominent example is Cut\&Count, which uses counting modulo two
to cancel disconnected candidates and obtain randomized
single-exponential algorithms parameterized by
treewidth~\cite{CNPPRW11}.
Other algorithms exploit restrictions on the supports of algebraic
computations: trimmed M\"obius inversion avoids evaluating irrelevant
parts of the subset lattice~\cite{BHKK08}, while maximum-degree-aware
sieving exploits zero evaluation points in inclusion--exclusion
formulas for counting occurrences of connected
patterns~\cite{BHKK18}.

\paragraph{Concurrent and independent work.}
Independently and concurrently, Kumabe~\cite{Kum26} obtained a deterministic $2^{O(k)}n^{O(1)}$-time algorithm for \textsc{Minimum Bisection} that also counts solutions. Both papers were submitted simultaneously to STOC 2027. The main dynamic programming ideas are similar. The difference between our approaches is that our decomposition guarantees \emph{subtree unbreakability} (property (D3) of \cref{thm:decomp}), which simplifies the dynamic program.

\section{Technical Overview}\label{sec:overview}

We first develop the main idea on a simpler problem, counting small sets with small cut, where the whole argument fits in a few paragraphs (\cref{sec:ov-warmup}). We then explain how an unbreakable decomposition turns \textsc{Minimum Bisection} into many instances of this problem, one per bag, and what changes there (\cref{sec:ov-bisection}).

\subsection{Warm-up: counting small cuts}\label{sec:ov-warmup}

Let $G$ be a graph on $n$ vertices, and let $s$ and $k$ be integers. For all $i\le s$ and $j\le k$, we want to count the sets $X\subseteq V(G)$ with $|X|=i$ and $|\delta(X)|=j$ in time $2^{O(s+k)}n^{O(1)}$. Listing all sets of size $s$ takes time $n^{s}$. Standard techniques such as color coding~\cite{AYZ95} find one such set within our time bound, but they do not count them.

We collect all sets into one polynomial. The \emph{monomial} of a set $X$ is $x^{|X|}z^{|\delta(X)|}$, and the \emph{partition function} of $W\subseteq V(G)$ is
\[
F(W)\;\coloneqq\;\sum_{X\subseteq W}x^{|X|}z^{|\delta(X)|},
\]
where the cut $|\delta(X)|$ is always taken in $G$, not in $G[W]$, and we discard every monomial with $x$-degree above $s$ or $z$-degree above $k$.\footnote{Formally, $F(W)$ lies in the ring $\Q[x,z]/(x^{s+1},z^{k+1})$. Discarding monomials is compatible with addition and multiplication, since exponents only grow.} The coefficient of $x^iz^j$ in $F(V(G))$ is the number we want. Computing $F(W)$ takes time $2^{|W|}n^{O(1)}$, so it is cheap for small $W$. Our goal is to compute $F(V(G))$.

Suppose that $W$ splits into two parts $W_1$ and $W_2$ with no edges between them. Then every $X\subseteq W$ splits into $X\cap W_1$ and $X\cap W_2$, and their cuts add up directly, since there are no edges between $W_1$ and $W_2$. This gives the following identity.
\[
F(W)\;=\;F(W_1)\,F(W_2).
\]

As will be seen later, this observation suggests that it will be convenient to take the logarithm, which turns the product into a sum, so that $\log F(W)=\log F(W_1)+\log F(W_2)$. Here $\log(1+f)=f-f^2/2+f^3/3-\cdots$ is the usual power series.\footnote{It is a finite sum: every monomial of $f=F(W)-1$ contains $x$, so $f^{s+1}=0$. The same holds for $\exp$, and the usual identities hold (\cref{lem:explog}).}

A natural way to separate the contributions to $\log F$ is to assign to each set $U$ the part not already accounted for by its proper subsets. Thus we define the \emph{contribution} $\Psi(U)$ recursively by $\Psi(U)=\log F(U)-\sum_{W\subsetneq U}\Psi(W)$.
This is M\"obius inversion on the subset lattice (see, e.g.,~\cite{BHKK08}); we state the identity in~\cref{lem:mobius}. Expanding this subtraction gives the explicit formula:
\[
\Psi(U)\;=\;\sum_{W\subseteq U}(-1)^{|U\setminus W|}\log F(W).
\]
By construction, $\log F(W)=\sum_{U\subseteq W}\Psi(U)$ for every $W$, and in particular
\begin{equation}\label{eq:ov-exp}
F(V(G))\;=\;\exp\Bigl(\sum_{U\subseteq V(G)}\Psi(U)\Bigr).
\end{equation}
By itself, \eqref{eq:ov-exp} saves nothing, since it sums over $2^n$ sets. The useful fact is that most contributions vanish. 
\begin{lemma}\label{lem:ov-candidates}
If $\Psi(U)\neq0$, then $|U|\le s$, $|\delta(U)|\le k$, and $G[U]$ is connected. We call sets satisfying these three conditions \emph{candidates}.
\end{lemma}
To see why, we temporarily give every vertex $v$ its own variable $x_v$ and every edge $e$ its own variable $z_e$, so that the monomial of a set $X$ becomes $\hat w(X)=\prod_{v\in X}x_v\prod_{e\in\delta(X)}z_e$, and the partition function of the whole graph becomes
\[
\hat F\;=\;\sum_{X\subseteq V(G)}\hat w(X) .
\]
Setting $x_v=x$ for $v\in W$, $x_v=0$ for $v\notin W$, and $z_e=z$ for every edge turns $\hat F$ into $F(W)$, and such substitutions commute with the logarithm.\footnote{Formally, we work with polynomials in the variables $x_v$ and $z_e$ in which every monomial of total degree above $s+k$ is discarded; such monomials would be mapped to $0$ anyway. The substitution is then a ring homomorphism, and ring homomorphisms commute with $\log$ (\cref{lem:explog}).} Setting $x_v=0$ for the vertices outside $W$ keeps exactly the monomials of $\log\hat F$ whose vertex variables lie in $W$. The next observation is that inclusion--exclusion over $W\subseteq U$ keeps those that use every vertex of $U$.

Consider some monomial that avoids some $w \in U$. Its contribution does not depend on $w$, but half of the sets $W \subseteq U$ contain $w$ and the other half do not. These sets have opposite signs in inclusion-exclusion. They contribute the same monomial in either case, since the subsets are the same, except for $w$, which does not feature in this monomial. Hence,
\begin{observation}\label{obs:ov-support}
$\Psi(U)$ is the sum of the monomials of $\log\hat F$ that use exactly the vertices of $U$, after substituting $x_v=x$ and $z_e=z$.
\end{observation}
We prove the analogous statement for bags in the proof of \cref{lem:candidates}, where monomials are grouped by the bag vertices they use.

\begin{proof}[Proof sketch of \cref{lem:ov-candidates}]
By \cref{obs:ov-support}, $\Psi(U)$ collects the monomials of $\log\hat F$ that use exactly the vertices of $U$. Since $\log\hat F=\sum_{r\ge1}(-1)^{r+1}(\hat F-1)^r/r$, every such monomial is a product of the monomials of nonempty sets $X_1,\dots,X_r$ whose union is $U$. Every edge leaving $U$ leaves some $X_i$, so the monomial contains a $x_v$ for every vertex of $U$ and a $z_e$ for every edge leaving $U$. Its $x$-degree is therefore at least $|U|$ and its $z$-degree at least $|\delta(U)|$. If $|U|>s$ or $|\delta(U)|>k$, the monomial is discarded. Finally, if $U=U_1\sqcup U_2$ with both parts nonempty and no edges between them, setting $x_v=0$ outside $U$ factorizes $\hat F$ into a part for $U_1$ and a part for $U_2$. Its logarithm is the sum of their logarithms, so no monomial uses vertices of both parts, and $\Psi(U)=0$.
\end{proof}

Therefore, it is enough to compute $\Psi(U)$ for each candidate $U \subseteq V(G)$, and we can easily compute $\log F(V(G)) = \sum_{U \subseteq V} \Psi(U)$, as $\Psi$ vanishes for non-candidates. This principle of eliminating disconnected contributions through logarithmic coefficients also underlies the algorithms of Patel and Regts for bounded-degree graphs~\cite{PR17,PR19}.

\paragraph{The algorithm.}
Checking all sets of size at most $s$ would take $n^{O(s)}$ time. But, the number of candidates is at most $n\cdot2^{s+k}$, and we can easily list them by branching as follows. For each vertex $v$, we list the candidates containing $v$ as follows. We grow a connected set $U\ni v$, and keep a set $D$ of edges that are declared cut. Start with $U=\{v\}$ and $D=\emptyset$. As long as some edge $e$ from $U$ to a vertex $w\notin U$ is not in $D$, branch: either add $w$ to $U$, if $|U|<s$, or add $e$ to $D$, if $|D|<k$. If neither action is permitted, terminate the branch without output. When no such edge is left, output $U$. Every branch increases $|U|+|D|$, which never exceeds $s+k$, so there are at most $2^{s+k}$ leaves. For completeness, let $U^\ast\ni v$ be a candidate, and follow the branch that adds $w$ if $w\in U^\ast$, and declares $e$ cut otherwise. This keeps $U\subseteq U^\ast$, and every edge in $D$ leaves $U^\ast$, so both budgets suffice, and the branch can only stop at $U=U^\ast$, because $U^\ast$ is connected.

So the algorithm is as follows. First, we list all the candidates and then compute the contribution $\Psi(U)$ of every candidate $U$ from the partition functions $F(W)$ with $W\subseteq U$. We return $F(V(G))=\exp(\sum_U\Psi(U))$, where $U$ ranges over the candidates. This is correct by~\eqref{eq:ov-exp} and \cref{lem:ov-candidates}. 
There are $\sum_{W\subseteq U}2^{|W|}=3^{|U|}\le3^s$ terms per candidate, and there are at most $n\cdot2^{s+k}$ candidates, so the running time is $2^{O(s+k)}n^{O(1)}$.

\subsection{Minimum Bisection}\label{sec:ov-bisection}

For \textsc{Minimum Bisection} we need the coefficient of $x^{\lfloor
n/2\rfloor}$, so $s=\lfloor n/2\rfloor$, and the warm-up does not apply
directly: the bound $|U|\le s$ is too expensive to list by branching, which takes $2^{O(s + k)}$ time.

To avoid that we use the decomposition of \cref{thm:decomp} as a black box.
Recall that in this decomposition, we have a rooted tree decomposition with
adhesions of size at most $2k$ (D1), in which every node is attached to its
parent's bag only through its adhesion (D2), and every bag $\beta(t)$ is
$(3k,k)$-edge-unbreakable in its subtree graph $G_t$ (D3). We construct it in
time $2^{O(k)}n^{O(1)}$ in \cref{sec:decomp}. We let $q=3k$ to denote the
unbreakability parameter.

Naturally, our algorithm is a dynamic programming over the decomposition. First,
we explain what each state of the dynamic program represents.  We call a partial
solution at a node $t$ a \emph{configuration}, that is a set $X\subseteq
\gamma(t)$ containing the vertices on side $A$; the other vertices of
$\gamma(t)$ are on side $B$. Node $t$ stores its \emph{table}, which counts its
configurations by their pattern $X\cap\sigma(t)$ on the adhesion, their size
$|X\cap \alpha(t)|$, and their cut:
\[
C_t(T,\ell,j)\;=\;\bigl|\{X\subseteq \gamma(t):\ X\cap\sigma(t)=T,\ |X\cap \alpha(t)|=\ell,\ |\delta_{G_t}(X)|=j\}\bigr|,
\]
for all $T\subseteq\sigma(t)$, $\ell\le n$ and $j\le k$. At the root,
$\sigma=\emptyset$ and $G_t=G$, so the table consists exactly of the numbers
$C_G(\ell,j)$ of \cref{thm:main}. In the transitions, we compute the table of
$t$ from the tables of its children. This is almost exactly the same as the
previous dynamic programs~\cite{CLPPS19,CKLPPSW21}. The only difference is that
they store the minimum cut instead of the number of configurations.

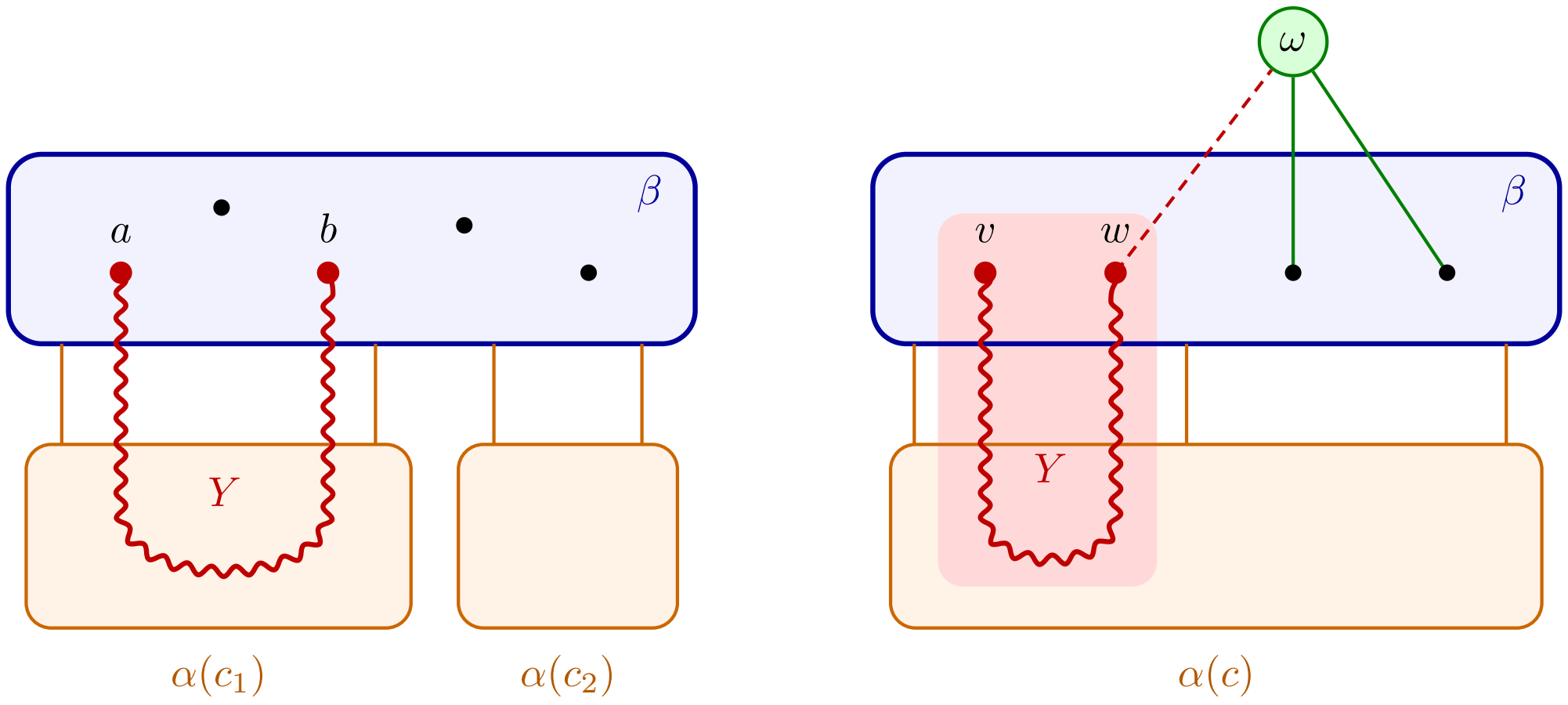
\begin{figure}[t]
\centering
\begin{tikzpicture}[
  font=\small,
  vtx/.style={circle,fill=black,inner sep=1.4pt},
  avtx/.style={circle,fill=red!75!black,inner sep=1.9pt},
  bag/.style={draw=blue!60!black,very thick,rounded corners=8pt},
  child/.style={draw=orange!80!black,thick,rounded corners=6pt},
  attachment/.style={draw=orange!80!black,thick},
  ypath/.style={draw=red!75!black,very thick,line cap=round,line join=round,decorate,decoration={snake,amplitude=1.2pt,segment length=6pt}},
  lab/.style={inner sep=2pt}]
  \begin{scope}
    \path[bag,fill=blue!5] (0,0) rectangle (5.8,1.6);
    \path[child,fill=orange!10] (0.15,-2.4) rectangle (3.4,-0.85);
    \path[child,fill=orange!10] (3.8,-2.4) rectangle (5.65,-0.85);
    \draw[attachment] (0.45,0)--(0.45,-0.85)
      (3.1,0)--(3.1,-0.85)
      (4.1,0)--(4.1,-0.85)
      (5.35,0)--(5.35,-0.85);
    \node[lab,blue!60!black,anchor=north east] at (5.6,1.5) {$\beta$};
    \node[lab,orange!70!black,anchor=north] at (1.775,-2.55) {$\alpha(c_1)$};
    \node[lab,orange!70!black,anchor=north] at (4.725,-2.55) {$\alpha(c_2)$};
    \coordinate (left-a) at (0.95,0.6);
    \coordinate (left-b) at (2.7,0.6);
    \draw[ypath] (left-a)--(0.95,-1.35)
      .. controls (0.95,-2.05) and (2.7,-2.05) .. (2.7,-1.35)
      --(left-b);
    \node[lab,red!75!black] at (1.825,-1.25) {$Y$};
    \node[avtx] at (left-a) {};
    \node[avtx] at (left-b) {};
    \node[lab,anchor=south] at (0.95,0.78) {$a$};
    \node[lab,anchor=south] at (2.7,0.78) {$b$};
    \node[vtx] at (1.8,1.15) {};
    \node[vtx] at (3.85,1.0) {};
    \node[vtx] at (4.9,0.6) {};
  \end{scope}
  \begin{scope}[xshift=7.3cm]
    \fill[blue!5,rounded corners=8pt] (0,0) rectangle (5.8,1.6);
    \fill[orange!10,rounded corners=6pt] (0.15,-2.4) rectangle (5.65,-0.85);
    \fill[red!15,rounded corners=6pt] (0.55,-2.05) rectangle (2.4,1.1);
    \draw[bag] (0,0) rectangle (5.8,1.6);
    \draw[child] (0.15,-2.4) rectangle (5.65,-0.85);
    \draw[attachment] (0.35,0)--(0.35,-0.85)
      (2.65,0)--(2.65,-0.85)
      (5.35,0)--(5.35,-0.85);
    \node[lab,blue!60!black,anchor=north east] at (5.6,1.5) {$\beta$};
    \node[lab,orange!70!black,anchor=north] at (2.9,-2.55) {$\alpha(c)$};
    \coordinate (right-v) at (0.95,0.6);
    \coordinate (right-w) at (2.05,0.6);
    \coordinate (right-b2) at (3.55,0.6);
    \coordinate (right-b3) at (4.85,0.6);
    \node[circle,draw=green!50!black,fill=green!15,thick,
      inner sep=3pt] (sink) at (3.55,2.55) {$\omega$};
    \draw[green!50!black,thick] (right-b2)--(sink) (right-b3)--(sink);
    \draw[red!75!black,thick,densely dashed] (right-w)--(sink);
    \draw[ypath] (right-v)--(0.95,-1.4)
      .. controls (0.95,-1.9) and (2.05,-1.9) .. (2.05,-1.4)
      --(right-w);
    \node[lab,red!75!black] at (1.5,-1.05) {$Y$};
    \node[avtx] at (right-v) {};
    \node[avtx] at (right-w) {};
    \node[vtx] at (right-b2) {};
    \node[vtx] at (right-b3) {};
    \node[lab,anchor=south] at (0.95,0.78) {$v$};
    \node[lab,anchor=south] at (2.05,0.78) {$w$};
  \end{scope}
\end{tikzpicture}
\caption{Left: a node with bag $\beta(t)$ and two children, whose components $\alpha(c_1)$ and $\alpha(c_2)$ (orange boxes) are attached to $\beta$ only through their adhesions (orange lines). Although $U=\{a,b\}$ is disconnected inside the bag, $U=Y\cap\beta(t)$ for a connected set $Y$ passing through a child component; hence $\Psi(U)$ may be nonzero. Right: listing with a sink. The sink $\omega$ is joined to every bag vertex except $v$. A connected set $Y\ni v$ (shaded) pays one edge to $\omega$ for each of its bag vertices other than $v$ (dashed), and nothing for its vertices inside the children.}
\label{fig:ov}
\end{figure}

\paragraph{Transitions of dynamic programming.} Fix a node $t$ with bag $\beta(t)$ ($\beta$ as a shorthand), write $H=G_t$, and suppose that the tables of the children are known. The goal is to compute the table of this node $t$. For $W\subseteq\beta$, the \emph{bag partition function} $F(W)$ counts the configurations whose bag part $X\cap\beta$ lies in $W$; child vertices are always allowed:
\[
F(W)\;=\sum_{X\subseteq \gamma(t):\ X\cap\beta\subseteq W}x^{|X\cap\beta|}\,y^{|X\cap \alpha(t)|}\,z^{|\delta(X)|}\,u^{X\cap\sigma(t)} .
\]
Here $x$ counts bag vertices, $y$ counts side-$A$ vertices outside the adhesion, $z$ counts the cut-size, and the markers $u^T=\prod_{v\in T}u_v$ record the pattern of the adhesion to the parent ($\sigma(t)$). Intuitively, this is very similar to the warmup where we kept track of only $x$ and $z$, but now we need adhesion information of $\sigma(t)$ to fill the table for each $T \subseteq \sigma(t)$, hence the markers, and we also need $y$ to track the size globally, whereas $x$ tracks the bag-size (which is what we will be able to assume is small due to unbreakability).

The next observation is that if $W$ is small,  we can easily compute $F(W)$ from the tables of the children. Fix the bag part $S=X\cap\beta$. Every edge of $H = G_t$ lies in the bag or in exactly one child. So the children are now ``independent'', and each child $c$ contributes one row of its table, the one for the pattern $S\cap\sigma(c)$. So $F(W)$ is a sum of $2^{|W|}$ products of table rows. As in the warm-up, $F(W)$ is cheap for small $W$, and the goal is $F(\beta)$, from which we read off the table of $t$. But $F(\beta)$ is a sum over all small sides $S\subseteq\beta$, and there are about $|\beta|^q$ of them. So we need to solve the case even when $\beta$ is large, using unbreakability. We will solve this large bag case exactly like in the warm-up.

\paragraph{Computing transitions in FPT time.} Without loss of generality, we count only the configurations with at most $q$ bag vertices on side $A$. By (D3), every other configuration of cut at most $k$ has at most $q$ bag vertices on side $B$, and we obtain those by complementing. So we discard every monomial with $x$-degree above $q$, besides those with $z$-degree above $k$. (Bags with at most $2q$ vertices we handle directly, by summing over all bag parts.)

To compute this restricted partition function, we define the contributions $\Psi(U)$, for $U\subseteq\beta$, exactly as in the warm-up: $\Psi(U)=\sum_{W\subseteq U}(-1)^{|U\setminus W|}\log F(W)$. 
\begin{lemma}[Analogue of~\cref{lem:ov-candidates}]\label{lem:ov-bag-candidates}
If $U\neq\emptyset$ and $\Psi(U)\neq0$, then $|U|\le q$, and $U=Y\cap\beta$ for a set $Y\subseteq\gamma(t)$ with $|\delta(Y)|\le k$ such that $H[Y]$ is connected.
\end{lemma}
So $U$ need not be connected in the bag, but we can still assume it is the restriction of a connected set $Y$ to the bag. $Y$ may connect its vertices through a child (\cref{fig:ov}, left). The proof uses the same monomial argument as in the warm-up; see \cref{lem:candidates} for the details.

\paragraph{Running time.} Similarly to the warmup, we call $U\subseteq\beta$ a
\emph{candidate} if $|U|\le q$ and $U=Y\cap\beta$ for a set $Y$ with $|\delta(Y)|\le
k$ such that $H[Y]$ is connected. As in the warm-up, every nonempty $U$ with
$\Psi(U)\neq0$ is a candidate, by \cref{lem:ov-bag-candidates}.  We cannot list
the candidates by branching as in the warm-up.  The set $Y$ may run deep into
the children, and branching might pay $2^{|Y|}$, while we can only afford to pay
$2^{|Y \cap \beta|}$, since that can be bounded by unbreakability. To solve this
technical issue, instead of branching, we enumerate candidates by using
important separators after attaching a sink.

For a vertex $v\in\beta$, add a new vertex $\omega$ joined by an edge to every bag vertex other than $v$. Let $Y\ni v$ satisfy $|Y\cap\beta|\le q$ and $|\delta_H(Y)|\le k$, with $H[Y]$ connected. In the new graph, $Y$ is still connected and its cut has size at most $k+q-1$: it contains the original cut edges and one new edge for each vertex of $(Y\cap\beta)\setminus\{v\}$. Its child vertices are free (\cref{fig:ov}, right). So a single $(v,\omega)$-cut of size at most $k+q-1$ pays both for the cut of $Y$ and for its bag vertices. By the theory of important cuts of Marx~\cite{Mar06}, some \emph{important} $(v,\omega)$-cut\footnote{An important $(v,\omega)$-cut is an inclusion-minimal such cut whose set of vertices reachable from $v$ cannot be strictly enlarged by a $(v,\omega)$-cut of at most the same size.} has size at most $k+q-1$ and a reachable set containing $Y$. This reachable set has at most $k+q-1$ bag vertices other than $v$, because each of them has its edge to $\omega$ in the cut. There are at most $4^{k+q}$ important $(v,\omega)$-cuts of size at most $k+q-1$~\cite{CLL09,CFKLMPPS15}. For each cut, list all subsets of its reachable bag vertices that contain $v$ and have size at most $q$. Each cut produces at most $2^{k+q}$ sets, so we list at most $8^{k+q}$ sets per vertex $v$, including all candidates containing $v$.

So the bag algorithm is as follows. First, we list a family containing every candidate and then compute the term $\Psi(U)$ for every listed set $U$ from the bag partition functions $F(W)$ with $W\subseteq U$. We also compute $\Psi(\emptyset)$ and obtain $F(\beta)=\exp(\sum_U\Psi(U))$, where $U$ ranges over the listed sets and the empty set. We recover the remaining configurations by complementing and read off the table of $t$. This is correct by \cref{lem:ov-bag-candidates} and the small-side argument above. There are $\sum_{W\subseteq U}2^{|W|}=3^{|U|}\le3^q$ terms per listed set, and we list at most $n\cdot8^{k+q}$ sets, so with $q=3k$ the running time is $2^{O(k)}n^{O(1)}$ per bag. The root table contains the answer.

\subsection{The decomposition.}
Finally, we give a short overview of the key cut-finding procedure in the decomposition construction, with the technical details deferred to \Cref{sec:decomp}. Based on~\cite{CLPPS19,ALLLS25} (see for example the technical overview of \cite{ALLLS25}), a simplified version of the main subroutine that we need to solve is this: Given a graph $G$, a terminal set $T \subseteq V(G)$ and a set $X \subseteq T$, find if there exists a set $Y\subseteq V(G)\setminus X$ with $|\delta(Y)|\leq i$ and $|Y\cap T|\geq i+1$ for some $i\leq k$. Given this cut-finding subroutine, a polynomial number of applications of an appropriate version of this subroutine essentially yields an $(O(k),k)$ edge-unbreakable decomposition. Previous approaches (for the stronger vertex unbreakability notion, \cite{CKLPPSW21, ALLLS25,Kor25}) use color coding or branching to detect such cuts. We use \emph{important} separators instead. 

The key observation is one which yields connectivity. Consider such a cut $\delta(Y)$. If $G[Y]$ is not connected, we can decompose it into connected components. The sum of the number of vertices of $T$ in these components is at least $i + 1$, and the total cut-size is at most $i$. It follows that there is one connected component $C$, such that the cut $\delta(V(C))$ has size at most $j$ and $|V(C) \cap T| \geq j + 1$ for some $j \leq i \leq k$. Thus, we can restrict our attention to cuts with $G[Y]$ connected. Note that this is where we use the fact that we are looking for edge-unbreakability: the analogous property is not true for vertex cuts. 

This leads to the main algorithm. Since our goal is to maximize the number of $T$-vertices in $Y$, we can push the cut as much towards $X$ as possible without increasing the cut-size, which is precisely why we can assume that the cut we consider is an important (edge) $(Y,X)$ separator, and hence, also a $(v,X)$ important separator for every $v \in Y$. So for every $v \in T \setminus X$, we simply enumerate all the $(v,X)$ important separators of size at most $k$ (there are at most $4^k$ of them for each $v$), and check if they satisfy the condition. This solves the problem in time $2^{O(k)}n^{O(1)}$, leading to the desired bound.  

\subsection{Organization}\label{sec:ov-org}

\Cref{sec:prelim} collects preliminaries: important cuts, truncated exponentials and logarithms, and the subset-sum identity used above. \Cref{sec:dp} gives the counting dynamic program in full detail, assuming the decomposition of \cref{thm:decomp}, and proves \cref{thm:main}. It also shows how to find a solution. \Cref{sec:decomp} constructs the decomposition and proves \cref{thm:decomp}. We conclude in \cref{sec:conclusion}. The warm-up is not used later: \cref{sec:dp} proves the candidate characterization for bags from scratch.

\section{Preliminaries}\label{sec:prelim}

All graphs are finite, simple and undirected. The input graph $G$ has $n$ vertices and $m$ edges. For a graph $H$ and $X\subseteq V(H)$, $\delta_H(X)$ is the set of edges of $H$ with exactly one endpoint in $X$, $|\delta_H(X)|$ is the \emph{cut size} of $X$, and $N_H(X)$ is the set of vertices outside $X$ with a neighbor in $X$. We omit the subscript when $H$ is clear from the context. For $Y\subseteq V(H)$, $H[Y]$ is the induced subgraph, $E(H[Y])$ is the set of edges with both endpoints in $Y$, and $H-E(H[Y])$ is the graph obtained from $H$ by deleting them. We use two facts about cuts. First, $|\delta(X)|=|\delta(V(H)\setminus X)|$. Second, if $X$ and $Y$ are disjoint and no edge joins them, then $\delta(X\cup Y)$ is the disjoint union of $\delta(X)$ and $\delta(Y)$; in particular, $|\delta(X)|$ is the sum of the cuts of the components of $H[X]$.

We use the notation of \cref{sec:intro-decomp}, and all bags of a tree decomposition are nonempty. The \emph{bag graph} of a node $t$ is $B_t=G[\beta(t)]-E(G[\sigma(t)])$.

A \emph{separation} of a graph $H$ is a pair $(A,B)$ with $A\cup B=V(H)$ and no edge between $A\setminus B$ and $B\setminus A$, and its \emph{order} is $|A\cap B|$. Let $A_0,B_0\subseteq V(H)$ be disjoint. By Menger's theorem, the minimum order of a separation $(A,B)$ with $A_0\subseteq A$ and $B_0\subseteq B$ equals the maximum number of vertex-disjoint paths from $A_0$ to $B_0$, and both can be computed in polynomial time. If $(A,B)$ has minimum order, then each path of a maximum family meets $A\cap B$ in exactly one vertex.

\paragraph{Important cuts.} Let $A_0,B_0\subseteq V(H)$ be disjoint and nonempty. An \emph{$(A_0,B_0)$-cut} is a set $F$ of edges such that $H-F$ has no path from $A_0$ to $B_0$, and $R_H(F)$ is the set of vertices reachable from $A_0$ in $H-F$. Every edge leaving $R_H(F)$ lies in $F$, so $|\delta(R_H(F))|\le|F|$. An $(A_0,B_0)$-cut is \emph{minimal} if no proper subset of it is an $(A_0,B_0)$-cut, and a minimal $(A_0,B_0)$-cut $F$ is \emph{important} if no $(A_0,B_0)$-cut $F'$ satisfies $|F'|\le|F|$ and $R_H(F)\subsetneq R_H(F')$.

\begin{lemma}[Important cuts~\cite{Mar06,CLL09,CFKLMPPS15}]\label{lem:important}
Let $H$ be a graph and $A_0,B_0\subseteq V(H)$ disjoint and nonempty.
\begin{enumerate}[(i)]
\item For every $(A_0,B_0)$-cut $F$ there is an important $(A_0,B_0)$-cut $F'$ with $|F'|\le|F|$ and $R_H(F)\subseteq R_H(F')$.
\item For every $p\ge0$, there are at most $4^p$ important $(A_0,B_0)$-cuts of size at most $p$, and they can be listed in time $O(4^p\,p\,(|V(H)|+|E(H)|))$.
\end{enumerate}
\end{lemma}

\paragraph{Truncated polynomial}
Throughout the paper we use truncated polynomials in finitely many variables with rational coefficients; these are polynomials that exclude certain monomials, together with all their multiples.
As an example, the exponent of a particular variable records a cut size; since we are only interested in cuts of size at most $k$, we exclude monomials where the corresponding exponent is larger than $k$.
Truncated polynomials naturally define a truncated polynomial ring where addition and multiplication are defined as usual, followed by discarding the excluded monomials.
Let $A$ be a truncated polynomial ring, and $V'$ be a subset of the variables.
Often, we want to focus only on monomials containing some variable from $V'$.
Let $J\subseteq A$ contain exactly the truncated polynomials whose every nonzero monomial contains at least one variable from $V'$ (and possibly other variables as well).
We call such a $J$ an ideal generated by $V'$. 
We always assume that \(J^{N+1}=0\) for some \(N\), meaning that every product of \(N+1\) elements of \(J\) is zero.

For a truncated polynomial $f$ satisfying $f^{N+1} = 0$ for some $N$, we set
\[
\exp(f)\;=\;\sum_{r=0}^{N}\frac{f^r}{r!},\qquad\log(1+f)\;=\;\sum_{r=1}^{N}(-1)^{r+1}\frac{f^r}{r}.
\]
These are the usual power series, all of whose omitted terms vanish (that is, these definitions do not depend on the choice of $N$).
Note that if $f\in J$ then $\exp(f)-1$ and $\log(1+f)$ lie in $J$. Throughout the paper, $\exp$ and $\log$ denote these finite series; we never apply them to numbers.
The following lemma shows that the usual identities relating exponentials and logarithms remain valid for truncated polynomials.
\begin{lemma}\label{lem:explog}
Let $A$ be a truncated polynomial ring and $J$ an ideal generated by a subset of its variables, with $J^{N+1}=0$ for some integer $N\ge0$. For $f,g\in J$:
\begin{enumerate}[(E1),leftmargin=*]
\item $\exp(\log(1+f))=1+f$.
\item $\log\bigl((1+f)(1+g)\bigr)=\log(1+f)+\log(1+g)$.
\item Let $A'$ be another truncated polynomial ring,
and let $h:A\to A'$ be a ring homomorphism
with $h(1)=1$. Then
$h(\log(1+f))=\log(1+h(f))$.
\end{enumerate}
\end{lemma}
\begin{proof}
Identities (E1) and (E2) follow from the standard formal power series
identities~\cite[Example~3.11 and Lemma~3.12]{Sam23} by substitution,
since $J^{N+1}=0$.
For (E3), $h$ preserves rational coefficients, sums, and products,
and hence the finite series defining $\log$.
\end{proof}

We use the standard M\"obius inversion formula on the subset
lattice~\cite[Section~2 of the conference version]{BHKK08}, applied coefficientwise to truncated polynomials.

\begin{lemma}[Möbius inversion]\label{lem:mobius}
Let $\beta$ be a finite set, let $R$ be a truncated
polynomial ring, and let $g,h:2^\beta\to R$.
Then $g(W)=\sum_{U\subseteq W}h(U)$ holds for every $W\subseteq\beta$
if and only if $h(U)=\sum_{W\subseteq U}(-1)^{|U\setminus W|}g(W)$
holds for every $U\subseteq\beta$.
\end{lemma}

\section{The Counting Dynamic Program}\label{sec:dp}
In this section we prove \cref{thm:main}, taking the decomposition of \cref{thm:decomp} as given. We follow the exposition of \cref{sec:ov-bisection}: we introduce tables, bag partition function and some basic results (\cref{sec:dp-intro}), listing candidates (\cref{sec:dp-candidates}), and the recursive step (\cref{sec:dp-recursive-step}). \cref{sec:proof-main} proves \cref{thm:main}.
Throughout \cref{sec:dp-intro,sec:dp-candidates,sec:dp-recursive-step}, the graph $G$ is connected, $k\ge1$, and $(\cT,\beta)$ is a rooted tree decomposition of $G$ with properties (D1)--(D3) of \cref{thm:decomp}. We write $q=3k$.
Here we also prove a basic result that shows how the subtree graph of a node splits into its bag graph and the subtree graphs of its children.

\begin{lemma}\label{lem:edgepartition}
Let $t$ have children $c_1,\dots,c_r$. Then
\begin{enumerate}[(i)]
\item The sets $\beta(t),\alpha(c_1),\dots,\alpha(c_r)$ partition $\gamma(t)$, and no edge joins distinct child components.
\item For each $i$, $N_G(\alpha(c_i))=\sigma(c_i)$ is a nonempty subset of $\beta(t)$.
\item $B_t,G_{c_1},\dots,G_{c_r}$ are subgraphs of $G_t$ with pairwise disjoint edge sets whose union is $E(G_t)$.
\end{enumerate}
\end{lemma}

\begin{proof}
(i) A vertex $w\in \alpha(c_i)$ lies in a bag below $c_i$ (including $\beta(c_i)$) but not in $\sigma(c_i)=\beta(c_i)\cap\beta(t)$, by definition of $\alpha(c_i)$.
Since the bags of a tree decomposition containing $w$ form a connected subtree, all of them lie below $c_i$. So $w\notin\beta(t)$, and $w$ lies in no other component. Every vertex of $\gamma(t)\setminus\beta(t)$ lies in a bag below some child $c_i$, and not in $\sigma(c_i)\subseteq\beta(t)$, so it lies in $\alpha(c_i)$. No bag contains vertices of two different components $\alpha(c_i),\alpha(c_j)$, so no edge joins them.

(ii) $N(\alpha(c_i))=\sigma(c_i)$ is (D2), and $\sigma(c_i)\subseteq\beta(t)$. If $\sigma(c_i)$ were empty, then $\alpha(c_i)$ would be a union of components of $G$, so $\alpha(c_i)=V(G)$ because $G$ is connected, contradicting $\beta(t)\neq\emptyset$, since $\alpha(c_i)$ and $\beta(t)$ are disjoint by (i).

(iii) Every edge of $G_{c_i}$ has an endpoint in $\alpha(c_i)$, which is disjoint from $\sigma(t)\subseteq\beta(t)$, so it is not inside $\sigma(t)$ and lies in $G_t$; clearly $B_t$ is a subgraph of $G_t$. An edge of $G_t$ with both endpoints in $\beta(t)$ lies in $B_t$, and in no $G_{c_i}$. Every other edge of $G_t$ has an endpoint in some $\alpha(c_i)$, and by (i) and (ii) its other endpoint lies in $\alpha(c_i)\cup\sigma(c_i)=\gamma(c_i)$. So it lies in $G_{c_i}$, and by (i) in no other child graph.
\end{proof}

\subsection{Tables and bag partition function}\label{sec:dp-intro}

A \emph{configuration} at a node $t$ is a set $X\subseteq \gamma(t)$. Its \emph{pattern} is $X\cap\sigma(t)$, its \emph{size} is $|X\cap \alpha(t)|$, and its \emph{cut} is $|\delta_{G_t}(X)|$. The \emph{table} of $t$ is
\[
C_t(T,\ell,j)\;=\;\bigl|\{X\subseteq \gamma(t):\ X\cap\sigma(t)=T,\ |X\cap \alpha(t)|=\ell,\ |\delta_{G_t}(X)|=j\}\bigr|
\]
for all $T\subseteq\sigma(t)$, $0\le\ell\le|\alpha(t)|$ and $0\le j\le k$.
That is, we are interested in sets with at most $k$ edges crossing them, and we also record their intersection with the adhesion $\sigma(t)$.
By (D1), this table has at most $4^k(n+1)(k+1)$ entries. At the root, $\sigma=\emptyset$, $\alpha(t)=V(G)$ and $G_t=G$, so $C_{\mathrm{root}}(\emptyset,\ell,j)=C_G(\ell,j)$.

Fix a node $t$ with children $c_1,\dots,c_r$, and write $\beta=\beta(t)$, $\sigma=\sigma(t)$, $H=G_t$ and $B=B_t$. Unless specified otherwise, cut sizes are taken in $H$. We work in the truncated polynomial ring
\[
R_t\;=\;\Q\bigl[x,y,z,(u_v)_{v\in\sigma}\bigr]\,\big/\,\bigl(x^{q+1},\ y^{|\alpha(t)|+1},\ z^{k+1},\ u_v^2\ (v\in\sigma)\bigr).
\]
The $x$ variable records the size of the intersection of a configuration with the bag (it is enough to only keep track of sizes up to $q$ because of the properties of the unbreakable decomposition), $y$ records the size, $z$ records the cut (we only care about cuts of size at most $k$), and the $u$ variables record the pattern.
In computational terms, an element of $R_t$ is an array of $(q+1)(|\alpha(t)|+1)(k+1)2^{|\sigma|}$ rationals, indexed by the exponents of $x,y,z$ and a set $T\subseteq\sigma$ of markers. Multiplication adds exponents and takes disjoint unions of marker sets, and it discards pairs of marker sets that intersect. For $T\subseteq\sigma$ we write $u^T=\prod_{v\in T}u_v$.

The \emph{monomial} of a configuration $X$, interpreted in $R_t$, is
\[
x^{|X\cap\beta|}\,y^{|X\cap \alpha(t)|}\,z^{|\delta(X)|}\,u^{X\cap\sigma}.
\]
In particular, the monomial of $X$ is zero if $|X\cap\beta| > q$ or $|\delta(X)|>k$.
The \emph{bag partition function} of $W\subseteq\beta$ is
\[
F(W)\;=\sum_{X\subseteq \gamma(t):\ X\cap\beta\subseteq W}x^{|X\cap\beta|}\,y^{|X\cap \alpha(t)|}\,z^{|\delta(X)|}\,u^{X\cap\sigma}.
\]
Intuitively, $F(W)$ counts the configurations $X$ with $X\cap\beta\subseteq W$, $|X\cap\beta|\le q$ and $|\delta(X)|\le k$, grouped by the size of their bag part, size, cut and pattern.

\begin{lemma}\label{lem:census}
For a child $c$ and $T\subseteq\sigma(c)$, let $P_c(T)=\sum_{\ell,j}C_c(T,\ell,j)\,y^\ell z^j$ be the row of the table of $c$ for the pattern $T$. Then for every $W\subseteq\beta$,
\[
F(W)\;=\;\sum_{S\subseteq W}x^{|S|}\,y^{|S\setminus\sigma|}\,z^{|\delta_B(S)|}\,u^{S\cap\sigma}\prod_{i=1}^{r}P_{c_i}(S\cap\sigma(c_i)).
\]
The same would hold if $R_t$ truncated $x^{q'+1}$ instead of $x^{q+1}$, for some $q'\ne q$.
Moreover, every monomial in $F(W)-1$ contains either $x$ or $z$.
\end{lemma}

\begin{proof}
This is the factorization of \cref{sec:ov-bisection}. By \cref{lem:edgepartition}(i), a configuration $X$ is the disjoint union of its bag part $S=X\cap\beta$ and the sets $X_{c_i}=X\cap \alpha(c_i)$. An edge of $B$ is cut by $X$ if and only if it is cut by $S$, and an edge of $G_{c_i}$ is cut by $X$ if and only if it is cut by $X\cap \gamma(c_i)=(S\cap\sigma(c_i))\cup X_{c_i}$. So by \cref{lem:edgepartition}(iii),
\[
|\delta_H(X)|\;=\;|\delta_B(S)|+\sum_{i=1}^{r}|\delta_{G_{c_i}}\bigl((S\cap\sigma(c_i))\cup X_{c_i}\bigr)|.
\]
Also $|X\cap \alpha(t)|=|S\setminus\sigma|+\sum_i|X_{c_i}|$ and $X\cap\sigma=S\cap\sigma$. So for a fixed $S$, the monomial of $X$ is $x^{|S|}y^{|S\setminus\sigma|}z^{|\delta_B(S)|}u^{S\cap\sigma}$ times a product over the children, in which the factor of $c_i$ depends only on $X_{c_i}$. The sets $X_{c_i}\subseteq \alpha(c_i)$ can be chosen independently, and by the definition of the table,
\[
\sum_{Y\subseteq \alpha(c_i)}y^{|Y|}z^{|\delta_{G_{c_i}}((S\cap \sigma(c_i))\cup Y)|}\;=\;P_{c_i}(S\cap \sigma(c_i))\qquad\text{in }R_t,
\]
since the terms with $z$-degree above $k$ vanish. This proves the formula. Notice that the proof is independent of $q$.

For the last claim, we show that $X=\emptyset$ is the only configuration whose monomial contains neither $x$ nor $z$. Let $X\neq\emptyset$ with $X\cap\beta=\emptyset$, and let $X$ intersect $\alpha(c)$. If $\alpha(c)\not\subseteq X$, then an edge of the connected graph $G[\alpha(c)]$ leaves $X$ (D2 of \Cref{thm:decomp}). Otherwise an edge joins $\alpha(c)$ to $\sigma(c)\neq\emptyset$ (\Cref{lem:edgepartition}), which avoids $X$. In both cases the edge lies in $H$, so $|\delta(X)|\ge1$.
\end{proof}

A configuration $X$ is \emph{small} if $|X\cap\beta|\le q$. For $T\subseteq\sigma$, $\ell \le |\alpha(t)|$ and $j\le k$, let $\Csm_t(T,\ell,j)$ be the number of small configurations with pattern $T$, size $\ell$ and cut $j$. It is the sum of the coefficients of $x^iy^\ell z^ju^T$ in $F(\beta)$ over $0\le i\le q$.

\begin{lemma}\label{lem:smallside}
If $|\beta|>2q$, then for all $T\subseteq\sigma$, $0\le\ell\le|\alpha(t)|$ and $0\le j\le k$,
\[
C_t(T,\ell,j)\;=\;\Csm_t(T,\ell,j)+\Csm_t\bigl(\sigma\setminus T,\ |\alpha(t)|-\ell,\ j\bigr).
\]
\end{lemma}

\begin{proof}
By (D3), every configuration $X$ with $|\delta(X)|\le k$ is small or satisfies $|\beta\setminus X|\le q$, and not both, since $|\beta|>2q$. The complement $X\mapsto \gamma(t)\setminus X$ preserves the cut, maps configurations of the second kind to small ones, maps the pattern $T$ to $\sigma\setminus T$, and maps the size $\ell$ to $|\alpha(t)|-\ell$.
\end{proof}

\subsection{Listing candidates}\label{sec:dp-candidates}

Let $J_t$ be the ideal generated by $\{x,z\}$. Every product of $q+k+1$ elements of $J_t$ has $x$-degree above $q$ or $z$-degree above $k$, so $J_t^{q+k+1}=0$.
By \cref{lem:census}, $\log F(W)$ is defined for every $W\subseteq\beta$, and it lies in $J_t$. The \emph{M\"obius term} of $U\subseteq\beta$ is
\[
\Psi(U)\;=\;\sum_{W\subseteq U}(-1)^{|U\setminus W|}\log F(W)\ \in J_t .
\]
By \cref{lem:mobius}, $\log F(W)=\sum_{U\subseteq W}\Psi(U)$ for every $W\subseteq\beta$. Taking $W=\beta$ and applying (E1), we get
\begin{equation}\label{eq:dp-exp}
F(\beta)\;=\;\exp\Bigl(\sum_{U\in\cU}\Psi(U)\Bigr)\qquad\text{for every family }\cU\supseteq\{U\subseteq\beta:\Psi(U)\neq0\}.
\end{equation}

The following lemma generalizes \cref{lem:ov-candidates} to bags.
\begin{lemma}\label{lem:candidates}
Let $U\subseteq\beta$ be nonempty with $\Psi(U)\neq0$. Then
\begin{enumerate}[(i)]
\item $|U|\le q$, and there is a set $Y\subseteq \gamma(t)$ with $Y\cap\beta=U$ such that
\item $|\delta(Y)|\le k$, and
\item $H[Y]$ is connected.
\end{enumerate}
\end{lemma}

\begin{proof}
As in \cref{sec:ov-warmup}, we now want to keep track of the exact vertices of a configuration, as well as the edges crossing it, rather than merely their number.
To this end, we give every vertex $v\in \gamma(t)$ a variable $x_v$ and every edge $e\in E(H)$ a variable $z_e$. Let $\Delta=q+|\alpha(t)|+k+|\sigma|$, and let $A$ be the ring of polynomials in these variables with rational coefficients, modulo all monomials of total degree above $\Delta$. The ideal $J_A$ generated by the variables satisfies $J_A^{\Delta+1}=0$. For $X\subseteq \gamma(t)$ let
\[
\hat w(X)\;=\;\prod_{v\in X}x_v\prod_{e\in\delta_H(X)}z_e,\qquad\text{and}\qquad\hat F\;=\;\sum_{X\subseteq \gamma(t)}\hat w(X)\ \in 1+J_A .
\]

\emph{Substitution.} For $W\subseteq\beta$, let $\varphi_W:A\to R_t$ be the substitution with $z_e\mapsto z$ for every edge, $x_v\mapsto y$ for $v\in \gamma(t)\setminus\beta$, and, for $v\in\beta$,
\[
x_v\;\mapsto\;\begin{cases}xy&\text{if }v\in W\setminus\sigma,\\ x\,u_v&\text{if }v\in W\cap\sigma,\\ 0&\text{if }v\notin W.\end{cases}
\]

Every variable of $A$ is mapped to zero or to a monomial
of positive total degree, so substitution cannot decrease the total
degree of a monomial with nonzero image. Since every nonzero monomial
of $R_t$ has total degree at most $\Delta$, every monomial
of $A$ of degree greater than $\Delta$ is mapped to zero.
Thus, $\varphi_W$ is a well-defined ring homomorphism.

By construction, $\varphi_W(\hat w(X))$ is the monomial of $X$ in $R_t$ if $X\cap\beta\subseteq W$, and is zero otherwise. So $\varphi_W(\hat F)=F(W)$, and by (E3),
\[
\varphi_W(\log\hat F)\;=\;\log F(W).
\]

As in \cref{obs:ov-support}, we group monomials by their support, now restricted to the bag. Write $\log\hat F=\sum_Mc_MM$, where $M$ ranges over the monomials of degree at most $\Delta$ and $c_M\in\Q$. The \emph{support} $\supp(M)$ of a monomial is the set of vertices $v$ such that $x_v$ divides $M$. If $\supp(M)\cap\beta\not\subseteq W$, then $\varphi_W(M)=0$, and otherwise $\varphi_W(M)=\varphi_\beta(M)$. Hence
\[
\log F(W)\;=\;\sum_{U\subseteq W}\Phi(U),\qquad\text{where}\quad\Phi(U)\;=\sum_{M:\ \supp(M)\cap\beta=U}c_M\,\varphi_\beta(M).
\]
By \cref{lem:mobius}, $\Psi(U)=\Phi(U)$ for every $U\subseteq\beta$.

Now let $U\neq\emptyset$ with $\Psi(U)\neq0$. Then some monomial $M$ with $\supp(M)\cap\beta=U$ has $c_M\neq0$ and $\varphi_\beta(M)\neq0$. Let $Y=\supp(M)$, so $Y\cap\beta=U$.

(i) Every $v\in U$ contributes a factor $x$ to $\varphi_\beta(M)$, so its $x$-degree is at least $|U|$. Since $\varphi_\beta(M)\neq0$, we get $|U|\le q$.

(ii) In $A$ we have $\log\hat F=\sum_{r=1}^{\Delta}(-1)^{r+1}(\hat F-1)^r/r$ by definitions of the logarithm and $A$.
Moreover, $(\hat F-1)^r$ is the sum of $\hat w(X_1)\cdots\hat w(X_r)$ over all tuples of nonempty sets $X_1,\dots,X_r\subseteq \gamma(t)$, by definition of $\hat F$ and the fact that $\hat w(\emptyset) = 1$. Since $c_M\neq0$, we have $M=\hat w(X_1)\cdots\hat w(X_r)$ for some such tuple. Then $Y=X_1\cup\dots\cup X_r$. Every edge $e\in\delta_H(Y)$ has an endpoint in some $X_i$ and the other endpoint outside $Y\supseteq X_i$, so $e\in\delta_H(X_i)$ and $z_e$ divides $M$. So the $z$-degree of $\varphi_\beta(M)$ is at least $|\delta(Y)|$, and $\varphi_\beta(M)\neq0$ gives $|\delta(Y)|\le k$.

(iii) Suppose that $Y=Y_1\sqcup Y_2$ with $Y_1,Y_2$ nonempty and no edge of $H$ between them. Let $\rho:A\to A$ be the substitution with $x_v\mapsto0$ for every $v\notin Y$, which keeps all other variables unchanged. For $X=X'\sqcup X''$ with $X'\subseteq Y_1$ and $X''\subseteq Y_2$, the set $\delta_H(X)$ is the disjoint union of $\delta_H(X')$ and $\delta_H(X'')$, so $\hat w(X)=\hat w(X')\hat w(X'')$. Hence
\[
\rho(\hat F)\;=\;\sum_{X\subseteq Y}\hat w(X)\;=\;\hat F_1\hat F_2,\qquad\text{where}\quad\hat F_i\;=\sum_{X\subseteq Y_i}\hat w(X),
\]
and by (E2) and (E3), $\rho(\log\hat F)=\log\hat F_1+\log\hat F_2$.

The left side is the sum of $c_{M'}M'$ over the monomials $M'$ with $\supp(M')\subseteq Y$, and every monomial on the right side has its support in $Y_1$ or in $Y_2$. Since $\supp(M)=Y$ intersects both, we get $c_M=0$, a contradiction. Therefore $H[Y]$ is connected.
\end{proof}

Following \cref{lem:candidates}, we generalize \cref{lem:ov-candidates}: a \emph{candidate} of $t$ is a set of the form $Y\cap\beta$, where $Y\subseteq \gamma(t)$, $|Y\cap\beta|\le q$, $|\delta(Y)|\le k$, and $H[Y]$ is connected.
Therefore, every nonempty $U$ with $\Psi(U)\neq0$ is a candidate.

\begin{lemma}\label{lem:listing}
Let $v\in\beta$. One can list, in time $O(8^{k+q}(k+q)(n+m))$, a family $\cU_v$ of at most $8^{k+q}$ subsets of $\beta$ that contains every candidate containing $v$.
\end{lemma}

\begin{proof}
This is the sink of \cref{sec:ov-bisection}. Let $\beta^-=\beta\setminus\{v\}$, and let $H^\omega$ be obtained from $H$ by adding a new vertex $\omega$ and an edge $w\omega$ for every $w\in\beta^-$. Let $Y\ni v$ be a set with $|Y\cap\beta|\le q$ and $|\delta_H(Y)|\le k$ such that $H[Y]$ is connected, and let $K_Y=\delta_{H^\omega}(Y)$. It consists of the at most $k$ edges of $\delta_H(Y)$ and the at most $q-1$ edges from $Y\cap\beta^-$ to $\omega$. So $K_Y$ is a $(\{v\},\{\omega\})$-cut of size at most $k+q-1$, and $R_{H^\omega}(K_Y)=Y$, because $H[Y]$ is connected. By \cref{lem:important}(i), some important $(\{v\},\{\omega\})$-cut $K$ satisfies $|K|\le k+q-1$ and $Y\subseteq R_{H^\omega}(K)$. Every vertex $w\in R_{H^\omega}(K)\cap\beta^-$ has its edge $w\omega$ in $K$, since $\omega$ is not reachable, so $|R_{H^\omega}(K)\cap\beta^-|\le k+q-1$.

So we list the at most $4^{k+q-1}$ important $(\{v\},\{\omega\})$-cuts $K$ of size at most $k+q-1$ (\cref{lem:important}(ii)), and for each of them we output the sets $\{v\}\cup U'$ for all $U'\subseteq R_{H^\omega}(K)\cap\beta^-$ with $|U'|\le q-1$. That is at most $2^{k+q-1}$ sets per cut, and at most $8^{k+q}$ sets in total. The output contains $Y\cap\beta=\{v\}\cup(Y\cap\beta^-)$, since $Y\cap\beta^-\subseteq R_{H^\omega}(K)\cap\beta^-$ and $|Y\cap\beta^-|\le q-1$. Listing the cuts takes time $O(4^{k+q}(k+q)(n+m))$, computing each reachable set $O(n+m)$, and writing each output set $O(q)$.
\end{proof}

\subsection{The recursive step}\label{sec:dp-recursive-step}

We now describe the recursive step, following \cref{sec:ov-bisection}; see \cref{alg:recursive-step}.

\begin{algobox}{Recursive step}\label{alg:recursive-step}%
\textbf{Input:} a node $t$ of the decomposition, and the tables $C_c$ of its children.\\
\textbf{Output:} the table $C_t$.
\begin{enumerate}[label=\arabic*:,ref=\arabic*,leftmargin=2.2em]
\item\label{line:small} \textbf{if} $|\beta|\le2q$ \textbf{then} compute $F(\beta)$ by \cref{lem:census}, but truncating $x^{|\beta|+1}$ instead of $x^{q+1}$, and \textbf{return} the sum of the coefficients of $x^iy^\ell z^ju^T$ in $F(\beta)$ over $0\le i\le |\beta|$, for all $T\subseteq\sigma$, $\ell$ and $j\le k$.
\item\label{line:list} $\cU\gets\{\emptyset\}\cup\bigcup_{v\in\beta}\cU_v$, with $\cU_v$ as in \cref{lem:listing}.
\item\label{line:psi} \textbf{for} every $U\in\cU$, compute $F(W)$ for all $W\subseteq U$ by \cref{lem:census}, and then $\Psi(U)$.
\item\label{line:exp} $F(\beta)\gets\exp\bigl(\sum_{U\in\cU}\Psi(U)\bigr)$.
\item\label{line:read} $\Csm_t(T,\ell,j)\gets$ the sum of the coefficients of $x^iy^\ell z^ju^T$ in $F(\beta)$ over $0\le i\le q$, for all $T\subseteq\sigma$, $\ell$ and $j\le k$.
\item\label{line:return} \textbf{return} $C_t(T,\ell,j)=\Csm_t(T,\ell,j)+\Csm_t(\sigma\setminus T,|\alpha(t)|-\ell,j)$ for all $T$, $\ell$ and $j$.
\end{enumerate}
\end{algobox}

\begin{lemma}\label{lem:recursive-step}
Given the tables of the children of $t$, \cref{alg:recursive-step} computes the table $C_t$ in time $2^{O(k)}n^{O(1)}$.
\end{lemma}

\begin{proof}
In Line~\ref{line:small} the bag size is small ($|\beta| \le 2q$) and we can thus straightforwardly compute $F(\beta)$ and from that $C_t$.
Otherwise $|\beta|>2q=6k$, so $k<n/6$. Every nonempty $U$ with $\Psi(U)\neq0$ is a candidate by \cref{lem:candidates}, so it lies in $\cU_v$ for every $v\in U$ by \cref{lem:listing}. So $\cU$ contains every $U$ with $\Psi(U)\neq0$, and Line~\ref{line:exp} computes $F(\beta)$ by \eqref{eq:dp-exp}. Lines~\ref{line:read} and~\ref{line:return} are correct by \cref{lem:smallside}.

For the running time, recall that $|\sigma|\le2k$ by (D1). 
An element of $R_t$ has at most $(q+1)(n+1)(k+1)2^{|\sigma|}$ coefficients, and a multiplication takes $O(3^{|\sigma|}((q+1)(n+1)(k+1))^2)$ operations, since there are $3^{|\sigma|}$ pairs of disjoint marker sets. 
By \cref{lem:census}, $F(W)$ is a sum of $2^{|W|}$ products of at most $O(n)$ factors. Line~\ref{line:small} sums $2^{|\beta|}\le4^q$ such products.
Line~\ref{line:list} takes time $O(n\cdot8^{k+q}(k+q)(n+m))$ by \cref{lem:listing}.
Every set in $\cU$ has at most $q$ elements, so Line~\ref{line:psi} needs $\sum_{W\subseteq U}2^{|W|}=3^{|U|}\le3^q$ products for each of the at most $1+n\cdot8^{k+q}$ sets $U\in\cU$, plus one logarithm for each $W\subseteq U$. Every $\exp$ and $\log$ takes $q+k$ multiplications, which also accounts for Line~\ref{line:exp}.
Line~\ref{line:read} and Line~\ref{line:return} straightforwardly take $2^{O(k)}n^{O(1)}$ operations.
With $q=3k$, this adds up to $2^{O(k)}n^{O(1)}$ operations in total.

We note that each operation only performs arithmetic between rationals with $n^{O(1)}$-bits representations, and can thus be performed in $n^{O(1)}$ time.
To see that, notice that the final entries of the tables as well as the coefficients of $F(\cdot)$ are integers bounded by $2^n$.
To compute $\Psi(U)$ we first compute the logarithm of $F(W)$ for some $W$; the denominators in the expression of the logarithm divide $(4k)!\le2^{O(n\log n)}$ and the numerators are at most $(2^n)^{O(k)}\le2^{O(n^2)}$, since $k<n/6$; therefore the sum of all these fractions can be represented by a pair (numerator, denominator) each of which needs $n^{O(1)}$ bits.
The same reasoning shows that $\Psi(U), \sum_{U\in \mathcal{U}}\Psi(U)$ and $\exp(\sum_{U\in \mathcal{U}}\Psi(U))$ can be represented as a pair of (numerator, denominator) each of which needs $n^{O(1)}$ bits.

We conclude that the total running time is $2^{O(k)}n^{O(1)}$.
\end{proof}

\subsection{Proof of the main theorem}\label{sec:proof-main}

\begin{proof}[Proof of \cref{thm:main}]
If $G$ is disconnected, we compute the numbers $C_{G_i}(\ell,j)$ for every component $G_i$ and combine them by convolution: $\sum_{\ell,j}C_G(\ell,j)\,y^\ell z^j$ is the product of the polynomials $\sum_{\ell,j}C_{G_i}(\ell,j)\,y^\ell z^j$, with all monomials of $z$-degree above $k$ discarded, because sizes and cuts add up over the components. If $G$ is connected and $k=0$, then $C_G(\ell,0)=1$ for $\ell\in\{0,n\}$ and $C_G(\ell,0)=0$ otherwise. So let $G$ be connected and $k\ge1$. We compute the decomposition of \cref{thm:decomp}, which has at most $n$ nodes, in time $2^{O(k)}n^{O(1)}$. Then we run \cref{alg:recursive-step} at every node, children before parents. By \cref{lem:recursive-step}, applied from the leaves to the root, every table is computed exactly, and the root table consists of the numbers $C_G(\ell,j)$. Each recursive step takes time $2^{O(k)}n^{O(1)}$.
\end{proof}

For the search problem, to find a bisection of size at most $k$, we use \cref{thm:main} as a black box, on graphs in which some vertices are forced to one side by large cliques.

\begin{lemma}\label{lem:forcing}
Let $G$ be a graph with $n$ vertices, let $P,Q\subseteq V(G)$ be disjoint, and let $a=n+k+2$ and $b=2n+k+3$. Let $G'$ be obtained from $G$ by adding a clique $A'$ on $a$ new vertices, each adjacent to every vertex of $P$, and a clique $B'$ on $b$ new vertices, each adjacent to every vertex of $Q$. Then, for all $0\le\ell\le n$ and $0\le j\le k$, $C_{G'}(a+\ell,j)$ is the number of sets $X\subseteq V(G)$ with $|X|=\ell$, $|\delta_G(X)|=j$, $P\subseteq X$ and $X\cap Q=\emptyset$.
\end{lemma}

\begin{proof}
Let $X'\subseteq V(G')$ with $|\delta_{G'}(X')|=j\le k$. Splitting a clique on $c$ vertices cuts at least $c-1$ of its edges, and $a-1,b-1>k$. So $X'$ contains all of $A'$ or none of it, and likewise for $B'$. If $A'\subseteq X'$ and some $v\in P$ is not in $X'$, then the $a>k$ edges between $v$ and $A'$ are cut. So $A'\subseteq X'$ implies $P\subseteq X'$. Similarly $A'\cap X'=\emptyset$ implies $P\cap X'=\emptyset$; the same holds for $B'$ and $Q$. Hence no edge inside $A'\cup B'$ or between $A'\cup B'$ and $V(G)$ is cut, and $|\delta_{G'}(X')|=|\delta_G(X'\cap V(G))|$. 

Finally, $|X'|$ is $|X'\cap V(G)|$, plus $a$ if $A'\subseteq X'$, plus $b$ if $B'\subseteq X'$. Since $|X'\cap V(G)|\le n<a$, $a+n<b$ and $b+n<a+b$, we have $|X'|=a+\ell$ with $0\le\ell\le n$ exactly when $A'\subseteq X'$, $B'\cap X'=\emptyset$ and $|X'\cap V(G)|=\ell$.
The result follows.
\end{proof}

\begin{proof}[Proof of \cref{cor:bisection}]
Let $h=\lfloor n/2\rfloor$. Every bisection has a side of size $h$, and if $n$ is even, both of its sides have size $h$. So the number of bisections of width $j\le k$ is $C_G(h,j)$ if $n$ is odd and $C_G(h,j)/2$ if $n$ is even, and the minimum width, if it is at most $k$, is the smallest $j$ with $C_G(h,j)>0$. To find a set $X$ with $|X|=h$ and $|\delta(X)|=j$, we maintain disjoint sets $P,Q$ such that some such $X$ satisfies $P\subseteq X$ and $X\cap Q=\emptyset$, starting with $P=Q=\emptyset$.

Let $C_{G,P,Q}(\ell,j)$ denote the number of sets $X\subseteq V(G)$ with $|X|=\ell, |\delta(X)|=j, P\subseteq X,$ and $X\cap Q=\emptyset$.
For an arbitrary vertex $v\notin P\cup Q$, let $P'=P\cup\{v\}$ and $Q'=Q\cup\{v\}$.
If $C_{G,P',Q}(h,j) > 0$ (which we compute using \Cref{lem:forcing} with $(P',Q)$) then we add $v$ to $P$.
Otherwise $C_{G,P,Q'}(h,j)>0$, as $C_{G,P',Q}(h,j) + C_{G,P,Q'}(h,j) = C_{G,P,Q}(h,j) > 0$; in this case we add $v$ to $Q$.
We proceed like that for all vertices, therefore in the end we obtain $X=P$.

This takes $n$ applications of \cref{thm:main} to graphs with $4n+2k+5$ vertices, so the running time is $2^{O(k)}n^{O(1)}$.
\end{proof}

\section{The Decomposition}\label{sec:decomp}

\decompositiontheorem*

In this section we prove \cref{thm:decomp}. Fix a connected graph $G$ and an integer $k\ge1$. The algorithm is recursive. A call $\textsc{Decompose}(D)$ receives a nonempty set $D\subseteq V(G)$ such that $G[D]$ is connected and $|N(D)|\le2k$, and returns a rooted tree decomposition of the graph
\[
H_D\;=\;G[D\cup N(D)]-E(G[N(D)])
\]
whose root bag contains $N(D)$. The whole algorithm is $\textsc{Decompose}(V(G))$, for which $N(D)=\emptyset$ and $H_D=G$. Inside a call we write $H=H_D$ and take cut sizes in $H$.

\begin{algobox}{Decompose}\label{alg:decompose}%
\textbf{Input:} a nonempty set $D\subseteq V(G)$ such that $G[D]$ is connected and $|N(D)|\le2k$.\\
\textbf{Output:} a rooted tree decomposition of $H=G[D\cup N(D)]-E(G[N(D)])$ whose root bag contains $N(D)$.
\begin{enumerate}[label=\arabic*:,ref=\arabic*,leftmargin=2.2em]
\item\label{line:H} $H\gets G[D\cup N(D)]-E(G[N(D)])$.
\item\label{line:base} \textbf{if} $|V(H)|\le2k+1$ \textbf{then return} a single node with bag $V(H)$.
\item\label{line:Z} $Z\gets N(D)\cup D_0$ for an arbitrary $D_0\subseteq D$ with $|N(D)\cup D_0|=2k+1$, and $\beta\gets V(H)$.
\item\label{line:while} \textbf{while} there exists a set $R=R_H(F)$, for $F$ (defined in~\cref{sec:decomp-alg}) with $\ex(R)\ge1$ \textbf{do}
\item\label{line:Z1} \quad $Z_1\gets$ any $|\delta(R)|+|R\cap Z|+1$ vertices of $(R\cap\beta)\setminus Z$.
\item\label{line:sep} \quad $(A,B)\gets$ a separation of $H$ minimizing $|A\cap B|$ subject to $Z_1\subseteq A$ and $Z\subseteq B$.
\item\label{line:refine} \quad $\beta\gets(\beta\cap B)\cup(A\cap B)$.
\item\label{line:recurse} Create a node $t$ with bag $\beta$, and for every component $D'$ of $H-\beta$, attach the tree returned by $\textsc{Decompose}(D')$ as a subtree below $t$.
\item\label{line:return-t} \textbf{return} the tree rooted at $t$.
\end{enumerate}
\end{algobox}

Skipping many details, the idea is as follows. We build the decomposition recursively, as in~\cite{CLPPS19,ALLLS25}: a call computes a root bag $\beta$ that contains $N(D)$ and is $(3k,k)$-edge-unbreakable in $H$, and recurses on the components of $H-\beta$, each of which has at most $2k$ neighbors, all in $\beta$. To compute the bag, we start with $\beta=V(H)$ and fix a \emph{reference set} $Z\subseteq\beta$ of $2k+1$ vertices that contains $N(D)$. A set that breaks $\beta$ has at most $k$ vertices of $Z$ on one side, and more than $3k$ vertices of $\beta$ on that side. Such a side has positive \emph{excess}: it contains more vertices of $\beta\setminus Z$ than it has cut edges and vertices of $Z$. Whenever we find a set of positive excess, we cut $\beta$ along a separation $(A,B)$ with $|A\cap B|\le2k$ between its vertices in $\beta\setminus Z$ and $Z$, and the bag becomes smaller. We search for such sets only among $2^{O(k)}n$ reachable sets of important cuts of size at most $k$. This suffices by an averaging argument (\cref{lem:averaging}): since the cut of a set is the sum of the cuts of its components, some component of a breaking set has positive excess on its own, and a reachable set that contains this component has positive excess as well. This argument keeps the running time single-exponential, and it is the step that needs edge cuts (\cref{rem:edges}).

\subsection{The algorithm behind \texorpdfstring{\cref{thm:decomp}}{Theorem 1.3}}\label{sec:decomp-alg}

A call maintains a bag $\beta$, initially all of $V(H)$, and a reference set $Z\subseteq\beta$ of $2k+1$ vertices that contains $N(D)$. Since $|Z|=2k+1$, for every $X\subseteq V(H)$, one of $X$ and $V(H)\setminus X$ contains at most $k$ vertices of $Z$. For $R\subseteq V(H)$, the \emph{excess} of $R$ is
\[
\ex(R)\;=\;|(R\cap\beta)\setminus Z|-|\delta(R)|-|R\cap Z|.
\]
In words, a positive excess means that $R$ contains more bag vertices outside $Z$ than it needs to pay for its cut and for its reference vertices. We consider the reachable sets $R_H(F)$ for
\begin{itemize}
\item every important $(A_0,Z\setminus A_0)$-cut $F$ in $H$ of size at most $k$, for every nonempty $A_0\subseteq Z$ with $|A_0|\le k$, and
\item every important $(\{v\},Z)$-cut $F$ in $H$ of size at most $k$, for every $v\in\beta\setminus Z$.
\end{itemize}
Each such set $R=R_H(F)$ contains the source of $F$ and avoids its sink, so $R\cap Z=A_0$ in the first case and $R\cap Z=\emptyset$ in the second. Hence every such set satisfies $|\delta(R)|\le|F|\le k$ and $|R\cap Z|\le k$. The excess and the reachable sets always refer to the current bag $\beta$.
Line~\ref{line:Z} is possible since $|V(H)|>2k+1$ after Line~\ref{line:base} and $|N(D)|\le2k$, and Line~\ref{line:Z1} is possible since $\ex(R)\ge1$.
This concludes the description of the algorithm.

Now, let us give a few properties behind the algorithm.
Throughout a call, $\beta$ denotes the current bag. We show that the loop maintains the following invariants.
\begin{enumerate}[label=(I\arabic*),leftmargin=*]
\item $Z\subseteq\beta$.
\item Every component $C$ of $H-\beta$ satisfies $|N_H(C)|\le2k$.
\end{enumerate}
Both hold after Line~\ref{line:Z}, where $H-\beta$ is empty.

\begin{lemma}\label{lem:smallsep}
In Line~\ref{line:sep}, $|A\cap B|\le |\delta(R)|+|R\cap Z|<|Z_1|$. In particular, $|A\cap B|\le2k$.
\end{lemma}

\begin{proof}
Let $B_R$ be the set of endpoints in $R$ of the edges of $\delta_H(R)$, so $|B_R|\le |\delta(R)|$. The pair
\[
\bigl(R,\;(V(H)\setminus R)\cup B_R\cup(R\cap Z)\bigr)
\]
is a separation of $H$: an edge from $R\setminus(B_R\cup Z)$ to $V(H)\setminus R$ would have its endpoint in $R$ in $B_R$. Its first side contains $Z_1\subseteq R$, its second side contains $Z$, and its separator has size $|B_R\cup(R\cap Z)|\le |\delta(R)|+|R\cap Z|=|Z_1|-1$. Since $(A,B)$ minimizes $|A\cap B|$, $|A\cap B|<|Z_1|$. Finally, $|\delta(R)|\le k$ and $|R\cap Z|\le k$ for every reachable set under consideration.
\end{proof}

\begin{lemma}\label{lem:refinement}
Lines~\ref{line:Z1}--\ref{line:refine} strictly decrease $|\beta|$ and preserve (I1) and (I2).
\end{lemma}

\begin{proof}
Let $\Gamma=A\cap B$ and $\beta'=(\beta\cap B)\cup\Gamma$.

\emph{The bag shrinks.} The vertices that leave the bag are those of $\beta\cap(A\setminus B)$, and those that enter it are those of $\Gamma\setminus\beta$. Since $Z_1\subseteq\beta\cap A$, we have $Z_1\setminus\Gamma\subseteq\beta\cap(A\setminus B)$. By \cref{lem:smallsep},
\[
|Z_1\setminus\Gamma|\;=\;|Z_1|-|Z_1\cap\Gamma|\;>\;|\Gamma|-|Z_1\cap\Gamma|\;\ge\;|\Gamma\setminus\beta|,
\]
where the last step uses $Z_1\cap\Gamma\subseteq\Gamma\cap\beta$. Hence $|\beta\cap(A\setminus B)|>|\Gamma\setminus\beta|$, and $|\beta'|<|\beta|$.

\emph{Invariant (I1).} Since $Z\subseteq B$ and $Z\subseteq\beta$, we have $Z\subseteq\beta\cap B\subseteq\beta'$.

\emph{Invariant (I2).} By Menger's theorem, there are $|\Gamma|$ vertex-disjoint paths from $Z_1$ to $Z$, and since $(A,B)$ minimizes $|A\cap B|$, each of them meets $\Gamma$ in exactly one vertex. For $x\in\Gamma$, let $\pi_x$ be the path through $x$, and let $\pi_x^-$ be its part before $x$. If $\pi_x^-$ is nonempty, it starts in $Z_1\subseteq A$, avoids $\Gamma$, and no edge joins $A\setminus B$ to $B\setminus A$, so $\pi_x^-\subseteq A\setminus B$.

Every vertex of $A\setminus B$ lies outside $\beta'\subseteq B$, and every vertex of $B\setminus A$ outside $\beta'$ lies outside $\beta$. Since no edge joins $A\setminus B$ to $B\setminus A$, every component $C'$ of $H-\beta'$ lies in $A\setminus B$ or in $(B\setminus A)\setminus\beta$. In the first case, the neighbors of $C'$ lie in $A$, and a neighbor in $A\setminus B$ would lie in $C'$. So $N(C')\subseteq\Gamma$, and $|N(C')|\le2k$ by \cref{lem:smallsep}. In the second case, $C'$ is connected and avoids $\beta$, so it lies in a component $C$ of $H-\beta$. Every neighbor of $C'$ lies in $B$ and in $\beta'$. A neighbor inside $C$ lies in $\beta'\setminus\beta\subseteq\Gamma$, and a neighbor outside $C$ lies in $N(C)\cap B$. So $|N(C')|\le|N(C)\cap B|+|\Gamma\cap C|$. For $x\in\Gamma\cap C$, the path $\pi_x$ starts in $Z_1\subseteq\beta$, outside $C$, and reaches $x\in C$. Let $y_x$ be the vertex of $\pi_x$ just before its first vertex in $C$. Then $y_x\in N(C)$ and $y_x\in\pi_x^-\subseteq A\setminus B$, and the vertices $y_x$ are distinct since the paths are disjoint. So $|\Gamma\cap C|\le|N(C)\cap(A\setminus B)|$, and $|N(C')|\le|N(C)|\le2k$ by (I2) for $\beta$.
\end{proof}

\begin{lemma}\label{lem:averaging}
Let $X\subseteq V(H)$ satisfy $|\delta(X)|\le k$, $|X\cap Z|\le k$ and $|X\cap\beta|\ge3k+1$. Then some reachable set $R$ under consideration satisfies $\ex(R)\ge1$.
\end{lemma}

\begin{proof}
Let $X_1,\dots,X_r$ be the vertex sets of the components of $H[X]$. Since $|\delta(X)|=\sum_i|\delta(X_i)|$, the excess is additive over them. By (I1), $|(X\cap\beta)\setminus Z|=|X\cap\beta|-|X\cap Z|$, and hence
\[
\sum_{i=1}^{r}\ex(X_i)\;=\;|(X\cap\beta)\setminus Z|-|\delta(X)|-|X\cap Z|\;\ge\;(3k+1-|X\cap Z|)-k-|X\cap Z|\;\ge\;1.
\]
So some $X_i$ has $\ex(X_i)\ge1$. Then $|\delta(X_i)|\le |\delta(X)|\le k$, $|X_i\cap Z|\le k$, and $(X_i\cap\beta)\setminus Z\neq\emptyset$. Let $A_0=X_i\cap Z$. If $A_0\neq\emptyset$, then $\delta(X_i)$ is an $(A_0,Z\setminus A_0)$-cut, and $Z\setminus A_0\neq\emptyset$ since $|Z|=2k+1>k$. If $A_0=\emptyset$, then $\delta(X_i)$ is a $(\{v\},Z)$-cut for any $v\in(X_i\cap\beta)\setminus Z$. In both cases the reachable set of this cut is $X_i$, because $H[X_i]$ is connected. By \cref{lem:important}(i), there is an important cut $F$ of the same kind with $|F|\le |\delta(X_i)|\le k$ and $R=R_H(F)\supseteq X_i$, and $R$ is one of the reachable sets under consideration, with $R\cap Z=A_0$. Also $|\delta(R)|\le|F|\le |\delta(X_i)|$ and $R\cap\beta\supseteq X_i\cap\beta$. Hence $\ex(R)\ge\ex(X_i)\ge1$.
\end{proof}

\begin{lemma}\label{lem:unbreakable}
When the loop of Line~\ref{line:while} ends, $\beta$ is $(3k,k)$-edge-unbreakable in $H$.
\end{lemma}

\begin{proof}
Let $X\subseteq V(H)$ with $|\delta(X)|\le k$. One of $X$ and $V(H)\setminus X$ contains at most $k$ vertices of $Z$, and both have the same cut. As the claim is symmetric in $X$ and its complement, we may assume that $|X\cap Z|\le k$. If $|X\cap\beta|\ge3k+1$, then \cref{lem:averaging} gives a reachable set of positive excess, so the loop would not have ended. So $|X\cap\beta|\le3k$.
\end{proof}

\subsection{Proof of \texorpdfstring{\cref{thm:decomp}}{Theorem 1.3}}\label{sec:decomp-global}

\begin{lemma}\label{lem:calls}
In Line~\ref{line:recurse}, every component $D'$ of $H-\beta$ satisfies $D'\subseteq D\setminus\beta$, $G[D']$ is connected, $N_G(D')=N_H(D')\subseteq\beta$, and $|N_G(D')|\le2k$. In particular, the recursive calls are legal.
\end{lemma}

\begin{proof}
By (I1), $N(D)\subseteq Z\subseteq\beta$, so $D'\subseteq D\setminus\beta$. The graph $H$ differs from $G[D\cup N(D)]$ only by edges inside $N(D)$, so $G[D']=H[D']$ is connected. Every neighbor in $G$ of a vertex of $D$ lies in $D\cup N(D)$, and an edge from $D'$ to $N(D)$ is not inside $N(D)$, so $N_G(D')=N_H(D')$. This set lies in $\beta$, since $D'$ is a component of $H-\beta$, and it has at most $2k$ vertices by (I2).
\end{proof}

\begin{lemma}\label{lem:global}
The call $\textsc{Decompose}(V(G))$ terminates and returns a rooted tree decomposition of $G$ with nonempty bags, at most $n$ nodes, and properties (D1)--(D3). Each call $\textsc{Decompose}(D)$ creates a node $t$ with $\gamma(t)=D\cup N(D)$, $\sigma(t)=N(D)$, and $G_t=H_D$.
\end{lemma}

\begin{proof}
For a call on $D$, let $t_D$ be the node it creates and $\beta_D$ its bag.

\emph{Bags.} We have $N(D)\subseteq\beta_D\subseteq D\cup N(D)$, and $\beta_D$ contains a vertex of $D$: in Line~\ref{line:base} because $D\neq\emptyset$, and otherwise because $\beta_D\supseteq Z$ by (I1) and $|Z|=2k+1>|N(D)|$.

\emph{Termination and size.} By \cref{lem:calls}, the sets of the child calls of $D$ are disjoint subsets of $D\setminus\beta_D$. So along every chain of nested calls the sets strictly decrease, and the recursion terminates. Moreover, the set $\beta_D\cap D$ is nonempty and disjoint from the sets of all calls below $D$, while calls that are not nested have disjoint sets. So the sets $\beta_D\cap D$ over all calls are pairwise disjoint and nonempty, and there are at most $n$ calls, and hence at most $n$ nodes.

\emph{Subtrees.} Let $t=t_D$. Every bag below $t$ has the form $\beta_{D''}\subseteq D''\cup N(D'')$ for a call $D''$ nested in $D$. Along the chain of calls from $D$ to $D''$, every call $D'$ made by a call $D_p$ satisfies $D'\subseteq D_p$ and $N(D')\subseteq\beta_{D_p}\subseteq D_p\cup N(D_p)$ by \cref{lem:calls}, so $D''\cup N(D'')\subseteq D\cup N(D)$. Conversely, let $w\in D\cup N(D)$ with $w\notin\beta_D$. Then $w$ lies in a component of $H_D-\beta_D$, that is, in the set of a child call. Following child calls whose sets contain $w$ as long as their bags do not contain $w$, we reach, since the chain is finite, a call whose bag contains $w$. Hence $\gamma(t)=D\cup N(D)$.

\emph{Adhesions.} Let $t'=t_{D'}$ be a child of $t=t_D$. Then $\beta(t')\subseteq D'\cup N(D')$, $D'\cap\beta(t)=\emptyset$, and $N(D')\subseteq\beta(t)\cap\beta(t')$ by \cref{lem:calls}. So $\sigma(t')=\beta(t')\cap\beta(t)=N(D')$. In particular, $\alpha(t')=\gamma(t')\setminus\sigma(t')=D'$, and $G_{t'}=G[D'\cup N(D')]-E(G[N(D')])=H_{D'}$. At the root, $D=V(G)$, $\sigma=\emptyset$ and $H_D=G$.

\emph{Properties (D1)--(D3).} Since $\sigma(t')=N(D')$ and $\alpha(t')=D'$, (D1) and (D2) follow from \cref{lem:calls}. Since $G_t=H_D$, (D3) follows from \cref{lem:unbreakable}; in Line~\ref{line:base}, $|\beta|\le2k+1$, so one side of every set has at most $k\le3k$ bag vertices.

\emph{Tree decomposition.} Every vertex lies in $\gamma(\mathrm{root})=V(G)$, and hence in some bag. Let $wv$ be an edge of $G$; it lies in $H_{V(G)}=G$. If $wv$ lies in $H_D$ but not both of its endpoints lie in $\beta_D$, then one endpoint lies in a component $D'$ of $H_D-\beta_D$, the other lies in $D'\cup N(D')$, and $wv$ is not inside $N(D')$, so $wv$ lies in $H_{D'}$. Following this chain, we reach a call whose bag contains both endpoints. Finally, fix a vertex $w$. The calls whose sets contain $w$ form a chain starting at the root, since sibling calls have disjoint sets; let $D^\ast$ be its last call. Every vertex of $H_{D^\ast}$ outside $\beta_{D^\ast}$ lies in the set of a child call, and $w\in D^\ast$ does not, so $w\in\beta_{D^\ast}$. Now let $t=t_D$ be any node whose bag contains $w$. If $w\notin D$, then $w\in N(D)=\sigma(t)$, so the bag of the parent of $t$ contains $w$ as well. So we can walk upwards from $t$ through nodes whose bags contain $w$ until we reach a node $t_{D'}$ with $w\in D'$; this happens at the latest at the root. There $w\in D'\cap\beta_{D'}$, so no child call of $D'$ contains $w$, and $D'=D^\ast$. So the nodes whose bags contain $w$ form a connected subtree.
\end{proof}

\begin{proof}[Proof of \cref{thm:decomp}]
By \cref{lem:global}, $\textsc{Decompose}(V(G))$ returns the required decomposition. It makes at most $n$ calls, and each call performs at most $n$ iterations of the loop by \cref{lem:refinement}, so the reachable sets are computed at most $n(n+1)$ times. Each time, there are fewer than $4^k+n$ sources, namely the $2^{2k}-1$ sets $A_0$ and the vertices of $\beta\setminus Z$, and each source has at most $4^k$ important cuts of size at most $k$, which are listed in time $O(4^kk(n+m))$ by \cref{lem:important}(ii). Evaluating the excess of a reachable set and computing a separation $(A,B)$ minimizing $|A\cap B|$ take polynomial time. So the running time is $2^{O(k)}n^{O(1)}$.
\end{proof}

\begin{remark}\label{rem:edges}
The construction needs edge cuts, rather than vertex separations, in one place: in \cref{lem:averaging}, the cut of $X$ is the sum of the cuts of the components of $H[X]$, so the excess is additive over the components. For a vertex separation $(A,B)$, the size of $A\cap B$ is not additive over components, because the components on one side may share separator vertices. For example, let $X$ be the set of leaves of a star whose center forms the separator. Then $X$ has many bag vertices and a separator of size $1$, but each leaf on its own also has a separator of size $1$, and hence excess $0$. The dynamic program uses edge cuts as well, in \cref{lem:listing}, where each bag vertex of $Y$ other than $v$ pays for its edge to the sink.
\end{remark}

\section{Conclusion}\label{sec:conclusion}

We gave a dynamic program over unbreakable decompositions that counts partial solutions instead of optimizing over them. Inside a bag, the logarithm of the partition function of partial solutions splits into M\"obius terms, each of which lives on the bag part of a connected set with at most $k$ cut edges, and important cuts towards a sink list these bag parts. Together with a single-exponential construction of decompositions whose bags are unbreakable in their subtrees, this counts the sets of every size with every cut up to $k$ in time $2^{O(k)}n^{O(1)}$, and in particular solves \textsc{Minimum Bisection} in this time.

Understanding the exact complexity of \textsc{Minimum Bisection} is an important
first step to better understand many graph problems. For example, Bliznets et
al.~\cite[Hypothesis~1.2 of the conference version]{BCKMP16} established conditional lower bounds,
based on a gap-hardness assumption for \textsc{Minimum Bisection}, for \textsc{Minimum Fill-In}, \textsc{Chain
Completion}, \textsc{Proper Interval Completion}, \textsc{Interval Completion},
\textsc{Threshold Completion}, and \textsc{Trivially Perfect
Completion}~\cite{BCKMP16}. We hope that adapting our algebraic counting approach will lead to faster
algorithms for these problems.

\paragraph{AI disclosure.}
Claude Opus-5.5 and GPT-6 Astra obtained the main improvement to the dynamic program. The idea for the decomposition is due to the authors. The plan for the entire presentation of this paper, including simplifying various proofs and the overview, is due to the authors. Opus and Astra assisted in writing.

\bibliographystyle{alpha}
\bibliography{min_bisection_single_exponential}

\end{document}